\newif\ifarxiv
\arxivtrue         

\ifarxiv
  \documentclass[sigconf,nonacm]{acmart}
  \makeatletter\gdef\@copyrightspace{}\makeatother
  \AtBeginDocument{\fancyhead{}}
\else
  \documentclass[sigconf,anonymous=false]{acmart} 
\fi

\usepackage{amsmath}
\usepackage{float}
\usepackage{graphicx}
\usepackage{xspace}
\DeclareRobustCommand{\RMone}{RM1\xspace}
\DeclareRobustCommand{\RMtwo}{RM2\xspace}
\usepackage{booktabs}
\usepackage{makecell}
\usepackage{multirow}
\usepackage{adjustbox} 
\usepackage{booktabs}
\usepackage{makecell}
\usepackage{subcaption} 
\usepackage{siunitx} 
\usepackage{etoolbox}
\usepackage{ifthen}
\usepackage[utf8]{inputenc}
\usepackage[ruled,vlined]{algorithm2e}
\usepackage{algpseudocode}
\SetKwInOut{Input}{Input}
\SetKwInOut{Output}{Output}
\SetKw{Let}{let}
\SetKw{Add}{add}
\SetKw{Return}{return}

\newenvironment{tightfloat}{%
  \begingroup
  \setlength{\intextsep}{6pt}%
  \setlength{\textfloatsep}{6pt}%
  \setlength{\abovecaptionskip}{3pt}%
  \setlength{\belowcaptionskip}{3pt}%
}{%
  \endgroup
}

\newboolean{makeXTRA}
\setboolean{makeXTRA}{false} 

\makeatletter%
\newboolean{maketitleAfterAllMeta}
\@ifclassloaded{acmart}%
{\setboolean{maketitleAfterAllMeta}{true}}%
{\setboolean{maketitleAfterAllMeta}{false}}%
\makeatother%

\usepackage{comment} 

\usepackage[utf8]{inputenc}
\usepackage[T1]{fontenc}

\usepackage{microtype}
 \usepackage{comment}
\usepackage{anyfontsize}
\usepackage[10pt]{moresize}

\usepackage{amsmath}
\usepackage{amsfonts}
\usepackage{mathtools}
\usepackage{float}

\makeatletter%
\@ifclassloaded{acmart}{}{\usepackage{amssymb}}
\makeatother%

\usepackage{siunitx}
\usepackage{xfrac}
\usepackage{tabu}
\usepackage{xspace}

\usepackage[normalem]{ulem}

\makeatletter%
\@ifclassloaded{acmart}{}%
{\usepackage[usenames,dvipsnames,table]{xcolor}}
\makeatother%

\makeatletter%
\@ifclassloaded{acmart}{}
{\@ifclassloaded{elsarticle}{}
{\usepackage{cite}}}
\makeatother%

\usepackage{setspace}

\usepackage{verbatim}

\usepackage{booktabs}
\usepackage{makecell}
\usepackage{multicol}
\usepackage{multirow}
\usepackage{tabularx}
\usepackage{array}

\usepackage{stfloats}

\makeatletter
\@ifclassloaded{IEEEtran}{%
  \let\MYcaption\@makecaption%
  \usepackage[font=footnotesize]{subcaption}%
  \let\@makecaption\MYcaption%
}{%
  \usepackage{subcaption}%
}%
\makeatother

\usepackage[textsize=tiny]{todonotes}

\makeatletter%
\@ifclassloaded{IEEEtran}{}{}%
\usepackage{enumitem}
\makeatother%

\usepackage{quoting}

\usepackage{listings}

\newcommand{\xspeed}[1]{\ensuremath{#1\times}}  

\lstdefinestyle{mystyle}{
    language=Python,           
    basicstyle=\ttfamily,      
    keywordstyle=\color{blue}, 
    frame=single,              
    breaklines=true,           
    captionpos=b               
}

\makeatletter%
\@ifclassloaded{acmart}%
{}{}%
\@ifclassloaded{elsarticle}%
{}{}
\@ifclassloaded{llncs}%
{}{}
\@ifclassloaded{jpconf}%
{%
  \usepackage{doi}%
}{}
\@ifclassloaded{IEEEtran}%
{%
  
  \newcommand{\ifIEEEtran}[1]{#1}
}%
{%
  \newcommand{\ifIEEEtran}[1]{}
  \usepackage{flushend}
}%
\makeatother%

\makeatletter%
\@ifpackageloaded{hyperref}%
{}%
{%
  \usepackage[hyphens]{url}
  \usepackage[breaklinks=true,bookmarks=false]{hyperref}
  \hypersetup{
    colorlinks=false,%
    pdfborder = 0 0 0
  }
}%
\makeatother%

\usepackage[capitalize,noabbrev,capitalize]{cleveref} 

\usepackage{stackengine}
\usepackage{graphicx}
\usepackage{tikz}
\usepackage{pgfplots}
\usepackage{pgfplotstable}

\ifthenelse{\boolean{makeXTRA}}
  {\newcommand{\XTRA}[1]{\phantom{}\begingroup\slshape\color{RoyalBlue}\ignorespaces#1\ignorespaces\endgroup}}
  {\newcommand{\XTRA}[1]{}}

\DeclareMathAlphabet{\mathpzc}{OT1}{pzc}{m}{it}

\makeatletter
\newcommand*{\textoverline}[1]{$\overline{\hbox{#1}}\m@th$}
\makeatother

\makeatletter
\@ifclassloaded{IEEEtran}{%
}{%
  
}%

\makeatletter 
\newcommand{\linebreakand}{%
  \end{@IEEEauthorhalign}
  \hfill\mbox{}\par
  \mbox{}\hfill\begin{@IEEEauthorhalign}
}
\makeatother 

\newlist{myReuseInterval}{enumerate}{1}
\setlist[myReuseInterval]{label=(R$_{\arabic*}$),nosep}

\crefname{section}{\S}{\S}
\crefformat{section}{#2\S#1#3}

\crefname{figure}{Fig.}{Figs.}
\Crefname{figure}{Figure}{Figures}

\crefname{equation}{Eq.}{Eqs.}

\definecolor{shadecolor}{gray}{0.9} 

\makeatletter
\@ifpackageloaded{algorithm2e}{\@ifclassloaded{IEEEtran}{%
  \SetAlFnt{\small}
  \SetAlCapFnt{\small}
  \SetAlCapNameFnt{\small}
}}%
\makeatother

\newcommand{\lstsetCommon}{%
  \lstset{%
    columns=fullflexible,%
    keepspaces=true,%
    escapeinside={<[}{]>},%
    moredelim=**[is][\color{Cerulean}]{<*}{*>},
    moredelim=**[is][\color{Red}]{<^}{^>},
    basicstyle=\ttfamily\footnotesize,%
    showstringspaces=false,%
    aboveskip=0em,%
    belowskip=0em,%
  }%
}

\newcommand{\lstsetMe}{%
  \lstsetCommon{}%
  \lstset{%
  }
}

\lstsetMe{}

\usepackage{soul}

\newcommand{\AcceptedManuscriptText}{%
  \textbf{Accepted manuscript (author’s version).}%
  \;This is the author’s version of the work, posted for your personal use.%
  \;Not for redistribution.%
  \;The definitive Version of Record will appear in \emph{SC Workshops ’25}.%
  \;DOI: \href{https://doi.org/10.1145/3731599.3767370}
  {10.1145/3731599.3767370}.%
}

\newcommand{\blfootnote}[1]{%
  \begingroup
  \renewcommand\thefootnote{}\footnote{#1}%
  \addtocounter{footnote}{-1}%
  \endgroup
}

\newif\ifhl{}
\hltrue{}
\hlfalse{}

\ifhl{}
 
\else
 
\fi

\newif\ifdraft{}
\drafttrue{}

\ifdraft{}
  \newcommand{\davidnote}[1]{ {\textcolor{purple} { ***[David]: #1 }}}
  \newcommand{\raynote}[1]{ {\textcolor{orange} { ***[Ray]: #1 }}}
  \newcommand{\ozgurnote}[1]{ {\textcolor{blue} { ***[Ozgur]: #1 }}}
  \newcommand{\NOTE}[1]{\phantom{}\begingroup\relax\ifmmode\boldmath\else\bfseries\fi\color{Cerulean}\ignorespaces#1\ignorespaces\endgroup}
  \newcommand{\TODO}[1]{\phantom{}\begingroup\relax\ifmmode\else\sffamily\fi\color{BurntOrange}\ignorespaces#1\ignorespaces\endgroup}
  \newcommand{\FIXME}[1]{\phantom{}\begingroup\relax\ifmmode\boldmath\else\bfseries\sffamily\fi\color{Red}\ignorespaces#1\ignorespaces\endgroup}
  \newcommand{\FIXED}[1]{\phantom{}\begingroup\relax\ifmmode\else\sffamily\fi\color{Green}\ignorespaces#1\ignorespaces\endgroup}
  \newcommand{\REPLACE}[1]{\phantom{}\begingroup\relax\ifmmode\else\sffamily\fi\color{Purple}\ignorespaces#1\ignorespaces\endgroup}
  \newcommand{\DELETE}[1]{\phantom{}\begingroup\relax\ifmmode\else\sffamily\fi\color{Red}\ifmmode\text{\sout{\ensuremath{#1}}}\else\sout{\ignorespaces#1\ignorespaces}\fi\endgroup}
\else
  \newcommand{\davidnote}[1]{}
  \newcommand{\raynote}[1]{}
  \newcommand{\ozgurnote}[1]{}
  \newcommand{\NOTE}[1]{}
  \newcommand{\TODO}[1]{}
  \newcommand{\FIXME}[1]{#1}
  \newcommand{\FIXED}[1]{#1}
  \newcommand{\DELETE}[1]{}
  \newcommand{\REPLACE}[1]{#1}
\fi

\def\BibTeX{{\rm B\kern-.05em{\sc i\kern-.025em b}\kern-.08em
    T\kern-.1667em\lower.7ex\hbox{E}\kern-.125emX}}

\ifarxiv
  \acmConference[]{}{}{}
  \acmBooktitle{}
  \acmYear{}
  \acmDOI{}
  \acmISBN{}
\else
    \copyrightyear{2025}
    \acmYear{2025}
    \setcctype{by}
    \acmConference[SC Workshops '25]{Workshops of the International Conference for High Performance Computing, Networking, Storage and Analysis}{November 16--21, 2025}{St Louis, MO, USA}
    \acmBooktitle{Workshops of the International Conference for High Performance Computing, Networking, Storage and Analysis (SC Workshops '25), November 16--21, 2025, St Louis, MO, USA}
    \acmDOI{10.1145/3731599.3767370}
    \acmISBN{979-8-4007-1871-7/2025/11}
\fi
\begin{document}
\title{Data Management System Analysis for Distributed Computing Workloads}

\begin{abstract}
Large-scale international collaborations such as ATLAS rely on globally distributed workflows and data management to process, move, and store vast volumes of data. ATLAS’s Production and Distributed Analysis (PanDA) workflow system and the Rucio data management system are each highly optimized for their respective design goals. However, operating them together at global scale exposes systemic inefficiencies, including underutilized resources, redundant or unnecessary transfers, and altered error distributions. Moreover, PanDA and Rucio currently lack shared performance awareness and coordinated, adaptive strategies.

This work charts a path toward co-optimizing the two systems by diagnosing data-management pitfalls and prioritizing end-to-end improvements. With the observation of spatially and temporally imbalanced transfer activities, we develop a metadata-matching algorithm that links PanDA jobs and Rucio datasets at the file level, yielding a complete, fine-grained view of data access and movement. Using this linkage, we identify anomalous transfer patterns that violate PanDA’s data-centric job-allocation principle. We then outline mitigation strategies for these patterns and highlight opportunities for tighter PanDA–Rucio coordination to improve resource utilization, reduce unnecessary data movement, and enhance overall system resilience.
\end{abstract}

\author{Kuan-Chieh Hsu}
\affiliation{%
  \institution{Brookhaven National Laboratory}
  \city{Upton}\state{NY}\country{USA}
}

\author{Sairam Sri Vatsavai}
\affiliation{%
  \institution{Brookhaven National Laboratory}
  \city{Upton}\state{NY}\country{USA}
}

\author{Ozgur O. Kilic}
\affiliation{%
  \institution{Brookhaven National Laboratory}
  \city{Upton}\state{NY}\country{USA}
}

\author{Tatiana Korchuganova}
\affiliation{%
  \institution{University of Pittsburgh}
  \city{Pittsburgh}\state{PA}\country{USA}
}

\author{Paul Nilsson}
\affiliation{%
  \institution{Brookhaven National Laboratory}
  \city{Upton}\state{NY}\country{USA}
}

\author{Sankha Dutta}
\affiliation{%
  \institution{Brookhaven National Laboratory}
  \city{Upton}\state{NY}\country{USA}
}

\author{Yihui Ren}
\affiliation{%
  \institution{Brookhaven National Laboratory}
  \city{Upton}\state{NY}\country{USA}
}

\author{David K. Park}
\affiliation{%
  \institution{Brookhaven National Laboratory}
  \city{Upton}\state{NY}\country{USA}
}

\author{Joseph Boudreau}
\affiliation{%
  \institution{University of Pittsburgh}
  \city{Pittsburgh}\state{PA}\country{USA}
}

\author{Tasnuva Chowdhury}
\affiliation{%
  \institution{Brookhaven National Laboratory}
  \city{Upton}\state{NY}\country{USA}
}

\author{Shengyu Feng}
\affiliation{%
  \institution{Carnegie Mellon University}
  \city{Pittsburgh}\state{PA}\country{USA}
}

\author{Raees Khan}
\affiliation{%
  \institution{University of Pittsburgh}
  \city{Pittsburgh}\state{PA}\country{USA}
}

\author{Jaehyung Kim}
\affiliation{%
  \institution{Carnegie Mellon University}
  \city{Pittsburgh}\state{PA}\country{USA}
}

\author{Scott Klasky}
\affiliation{%
  \institution{Oak Ridge National Laboratory}
  \city{Oak Ridge}\state{TN}\country{USA}
}

\author{Tadashi Maeno}
\affiliation{%
  \institution{Brookhaven National Laboratory}
  \city{Upton}\state{NY}\country{USA}
}

\author{Verena Ingrid Martinez Outschoorn}
\affiliation{%
  \institution{University of Massachusetts}
  \city{Amherst}\state{MA}\country{USA}
}

\author{Norbert Podhorszki}
\affiliation{%
  \institution{Oak Ridge National Laboratory}
  \city{Oak Ridge}\state{TN}\country{USA}
}

\author{Frédéric Suter}
\affiliation{%
  \institution{Oak Ridge National Laboratory}
  \city{Oak Ridge}\state{TN}\country{USA}
}

\author{Wei Yang}
\affiliation{%
  \institution{SLAC National Accelerator Laboratory}
  \city{Menlo Park}\state{CA}\country{USA}
}

\author{Yiming Yang}
\affiliation{%
  \institution{Carnegie Mellon University}
  \city{Pittsburgh}\state{PA}\country{USA}
}

\author{Shinjae Yoo}
\affiliation{%
  \institution{Brookhaven National Laboratory}
  \city{Upton}\state{NY}\country{USA}
}

\author{Alexei Klimentov}
\affiliation{%
  \institution{Brookhaven National Laboratory}
  \city{Upton}\state{NY}\country{USA}
}

\author{Adolfy Hoisie}
\affiliation{%
  \institution{Brookhaven National Laboratory}
  \city{Upton}\state{NY}\country{USA}
}

\maketitle
\ifarxiv
  \blfootnote{\AcceptedManuscriptText}

  \fancyfoot{}%
  \fancyfoot[C]{\footnotesize Accepted manuscript — Version of Record: %
    \href{https://doi.org/10.1145/3731599.3767370}{10.1145/3731599.3767370}}
\fi




\vspace{-0.2cm}
\section{Introduction}
\label{sec:intro}
\vspace{-0.1cm}
Large-scale scientific experiments increasingly depend on distributed computing infrastructures for processing and analyzing the vast amounts of data they generate, where workflow and data management systems must operate in tandem. In the ATLAS (A Toroidal LHC Apparatus) experiment~\cite{atlas}, two such systems, the Production and Distributed Analysis (PanDA) workload manager~\cite{panda} and the Rucio data management system~\cite{rucio}, form the backbone of distributed analysis. Each system has been optimized for its individual objectives. PanDA is responsible for scheduling and monitoring millions of jobs, while Rucio manages the placement and transfer of petabytes of data. However, it remains unclear whether these systems achieve collective efficiency as a unified architecture when deployed across globally distributed resources. The central question motivating this work is \textit{whether file transfers triggered by job execution are performed efficiently, and whether systematic anomalies reveal coordination inefficiencies that compromise system performance}.

Answering this question requires correlating past job executions with the file transfers that supported them. However, such an analysis faces several challenges. (1) \textbf{Lack of direct metadata mapping.} PanDA and Rucio metadata are not directly linked. Transfer records do not carry job identifiers, which makes it difficult to trace file movements back to the jobs that required them. (2) \textbf{Scale of operations.} The sheer scale of ATLAS operations is daunting. PanDA executes millions of jobs per week, while Rucio orchestrates tens of millions of individual file transfers, producing metadata that is massive in both volume and complexity. (3) \textbf{Metadata quality.} The metadata itself is often heterogeneous and incomplete, with issues such as missing site information, inconsistent file attributes, or incomplete records - all of which hinder strict correlation. (4) \textbf{Divergent optimization goals.} PanDA prioritizes job throughput, while Rucio emphasizes balanced data distribution. These differing objectives can lead to conflicting behaviors such as redundant transfers, overloaded sites, or altered error patterns. (5) \textbf{Anomaly detection complexity.} Detecting anomalies at this scale requires linking diverse metadata sources and distinguishing true inefficiencies from routine variability in system behavior.

To address these challenges, we conduct an extensive data analysis across both PanDA and Rucio by integrating metadata from millions of jobs and file transfers. Our analysis provides a unified view of workflow execution and data movement. We bridge the metadata gaps through individual file mapping, scale correlation to massive datasets, handle incomplete records, and uncover systemic inefficiencies resulting from uncoordinated optimization. While our analysis focuses on ATLAS infrastructure, the methodology and insights are applicable to other large-scale scientific experiments and distributed computing environments that face similar coordination challenges between workflow and data management systems.

Our key contributions in this paper are summarized below:
\begin{itemize}
    \item We developed a fine-grained matching algorithm that links PanDA jobs with Rucio file transfers at the file level, overcoming the absence of direct identifiers and enabling systematic joint analysis of the two systems.
    \item Using this framework, we analyzed millions of jobs and millions of transfers, providing a comprehensive characterization of workflow–data interactions at ATLAS scale.
    \item We introduced relaxed matching strategies that handle incomplete metadata, expanding analysis coverage and enabling anomaly detection beyond exact matching.
    \item Our analysis uncovered anomalous behaviors including redundant transfers, prolonged staging delays, bandwidth under-utilization, and site-level imbalances that degrade performance despite local optimizations.
    \item We presented detailed case studies that illustrate how inefficiencies manifest in practice, and we discuss opportunities for tighter co-design between PanDA and Rucio to improve efficiency, resource utilization, and system resilience.
\end{itemize}
\vspace{-0.1cm}
The remainder of this paper is structured as follows. Section~\ref{sec:background} provides essential background on the ATLAS computing infrastructure, with particular emphasis on the PanDA and Rucio systems. Section~\ref{sec:motivation} motivates the present study by analyzing existing limitations and performance challenges. Section~\ref{sec:dataanalysis} outlines the proposed methodology, including the exact mapping algorithm and alternative relaxed matching strategies. Section~\ref{sec:experiments} reports the results of the analysis, supplemented by case studies that illustrate notable performance anomalies. Section~\ref{sec:related} discusses the related work, and Section~\ref{sec:conclusion} concludes by discussing implications for future system design and optimization.

\vspace{-0.2cm}
\section{Background}
\label{sec:background}
\begin{figure}[h!]
\vspace{-0.2cm}
  \raggedright
  \hspace*{-0.04\linewidth}
    \includegraphics[width=1.05\linewidth]{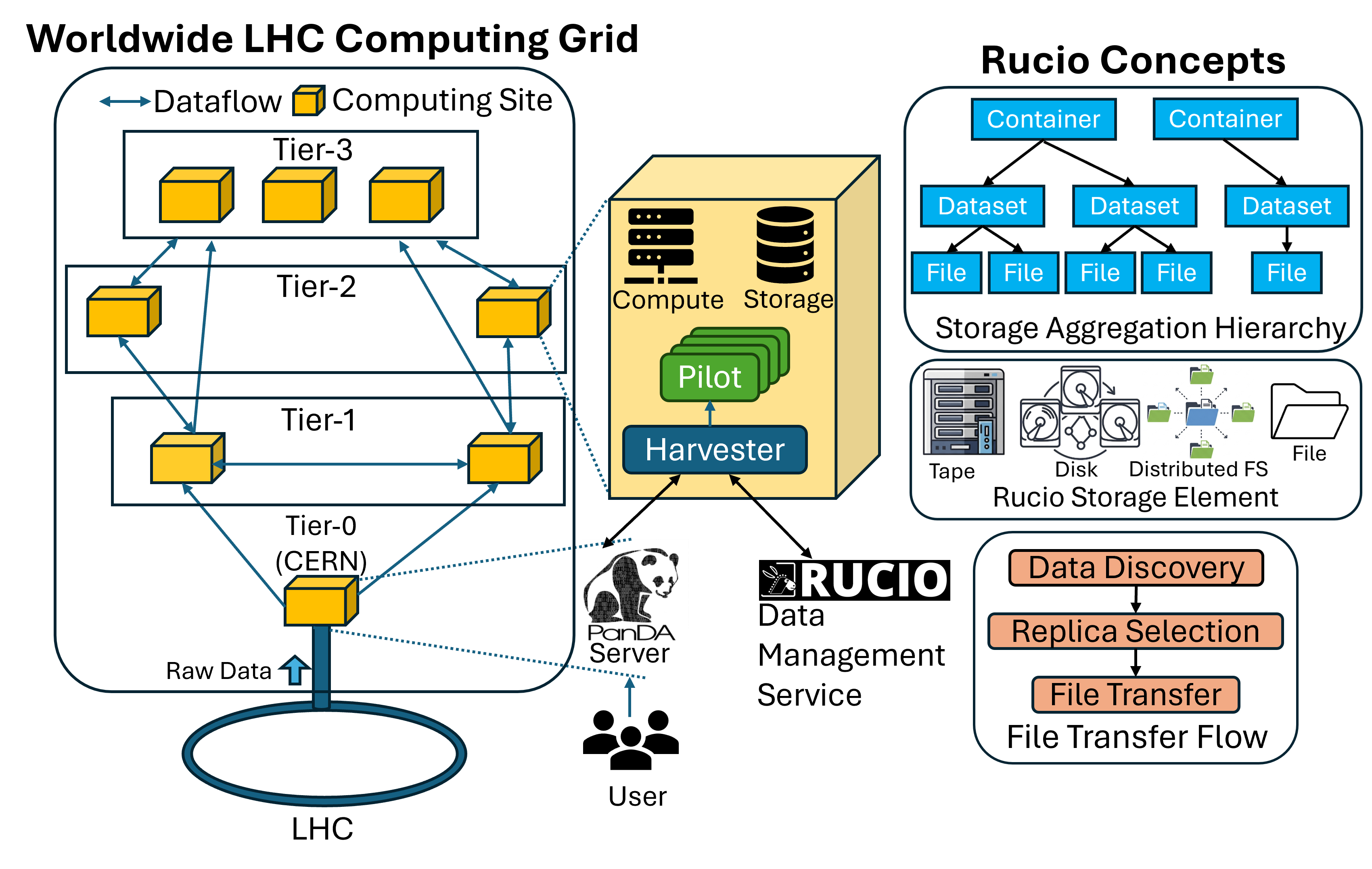}
    \caption{Architectural flow of the ATLAS workload and data management systems.}
    \label{fig:panda_rucio_flow}
    \vspace{-0.6cm}
\end{figure}

\vspace{-0.1cm}
ATLAS is a high-energy physics experiment at the Large Hadron Collider (LHC) located at the European Organization for Nuclear Research (CERN) in Geneva, Switzerland. The ATLAS detector investigates particle collisions at high energies, generating petabytes of data annually in the search for new physics discoveries. Thousands of scientists across the world analyze this massive dataset remotely using the ATLAS globally distributed computing infrastructure which spans approximately 200 computing centers across more than 40 countries. This distributed analysis ecosystem relies on two critical systems: PanDA for workload management and Rucio for data management, which together coordinate the complex computational demands of modern particle physics research. Figure~\ref{fig:panda_rucio_flow} illustrates the architectural flow of these two systems across the Worldwide LHC Computing Grid (WLCG). In the following, we provide a brief overview of these systems and their operation. 

\vspace{-0.2cm}
\subsection{PanDA System}
\vspace{-0.1cm}
The PanDA system is ATLAS's workload management system, responsible for scheduling, executing, and monitoring large-scale analysis and simulation jobs across the WLCG. Figure \ref{fig:panda_rucio_flow} shows the categorization of WLCG computing sites, which are organized into four tiers.
\textbf{Tier-0:} Located at CERN, the Tier-0 records raw detector data and performs the initial processing, storing results on large-scale tape and disk systems.
\textbf{Tier-1:} These sites are large national laboratories connected by high-capacity networks. They provide long-term storage and perform large-scale data reprocessing.
\textbf{Tier-2:} Typically medium-sized universities and laboratories, Tier-2 sites are linked to Tier-1 centers and contribute storage and computing resources for simulation and user analysis.
\textbf{Tier-3:} These are smaller institutions with limited resources that usually support localized data access for individual researchers.

At the center of this ecosystem is the PanDA server, located at Tier-0. It receives jobs submitted by users, who specify their input and output datasets. Jobs are placed in a global queue and assigned to computing sites by a brokerage module, based on many criteria such as job type, priority, input data location, and site availability. 

At each site, PanDA interacts with the Harvester service, which orchestrates execution by deploying lightweight Pilot jobs to worker nodes. Pilots provision the execution environment, validate resources, and then request a payload job from the dispatcher, thereby shielding workload jobs from grid heterogeneity. In addition to managing pilots, Harvester communicates with the Rucio data management system for dataset discovery, transfers, and output registration. This ensures that input data are staged at the site storage before execution and that outputs are properly integrated into the global dataset catalog.

\vspace{-0.2cm}
\subsection{Rucio System}
\vspace{-0.1cm}
Rucio is ATLAS’s distributed data management system, operating in close integration with PanDA to ensure that jobs and their required data are properly co-located or network-accessible across the WLCG. Figure~\ref{fig:panda_rucio_flow} illustrates Rucio's main concepts, including data hierarchy, storage abstraction, and transfer workflow. Rucio employs a three-tiered namespace hierarchy of data. The smallest unit is the file, which is grouped into datasets to enable bulk operations such as transfers or deletions. Multiple datasets may be aggregated into containers, which can themselves be nested, enabling flexible grouping of large-scale collections such as raw detector streams. All data are referenced by globally unique Data Identifiers (DIDs), ensuring immutable naming and provenance. 

In Rucio, physical data locations are abstracted through Rucio Storage Elements (RSEs), which represent logical endpoints for disks, tapes, cloud storage, or distributed file systems. A key concept in Rucio is the replica, denoting a physical copy of a file stored at a particular site. Since ATLAS data are globally distributed, the same file may have replicas at multiple RSEs. To manage data placement, Rucio employs replication rules, which specify where data must exist, how many replicas must be maintained, and the duration of retention. When rules are applied to a DID, Rucio automatically triggers the transfer of missing replicas and protects them from deletion until all rules expire.

\begin{figure}[t]
    \centering
    \vspace{-1.2cm}
    \includegraphics[width=1\linewidth]{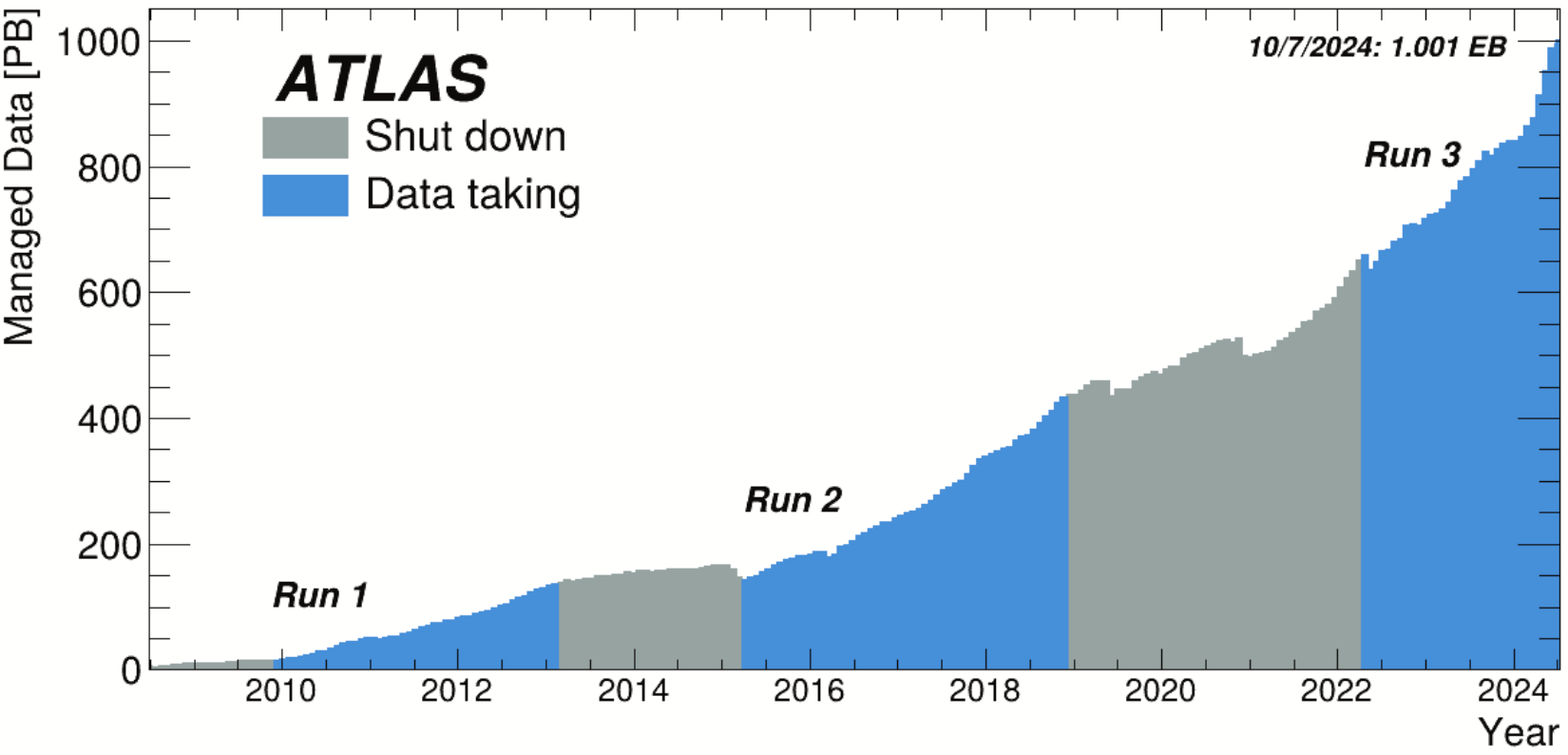}
    \vspace{-1.7cm}
    \caption{Total ATLAS volume managed by Rucio, approaching 1 exabyte of data in mid-2024~\cite{park2024ai}.}
    \label{fig:datavol}
    \vspace{-0.5cm}
\end{figure}

For job execution, PanDA and Harvester coordinate with Rucio. When PanDA assigns a job to a site, Harvester queries Rucio to resolve dataset locations. If replicas are not available locally, Rucio evaluates replication rules, creates new replicas by transferring files to the site's RSE, and confirms accessibility before execution begins. The Rucio file transfer workflow includes three steps as shown in Figure \ref{fig:panda_rucio_flow}: (1) data discovery, to determine whether required datasets already exist at the destination, (2) replica selection, to choose the best source replica based on protocol, throughput, and network performance metrics, and (3) file transfer. The scale at which this workflow operates becomes evident when examining the data volumes managed by Rucio over time. Figure \ref{fig:datavol} illustrates the cumulative growth of ATLAS data managed by Rucio from 2009 to 2024. By mid-2024, Rucio managed over 1 exabyte of ATLAS data, representing more than a doubling of the data volume since 2018. This massive dataset is distributed across more than 180 storage elements worldwide \cite{rucio}, while maintaining high availability and performance for both automated workflows and interactive user access.

\vspace{-0.2cm}
\section{Motivation}
\label{sec:motivation}
\vspace{-0.1cm}
While PanDA and Rucio each fulfill their roles effectively within ATLAS’s distributed computing environment, their independent design objectives raise important questions about system-level efficiency and resilience. When both are deployed together across a vast and heterogeneous grid, mismatches between their optimization principles can lead to unforeseen inefficiencies. Our investigation is motivated by two key goals: (1) examining how disjointed optimizations between the two systems may compromise overall performance, and (2) analyzing the systemic vulnerabilities and resilience challenges that arise from large-scale, imbalanced data movement.

\vspace{-0.2cm}
\subsection{Disjointed Workload Management and Data Management Optimizations}
\vspace{-0.1cm}
Because of the vast volume of scientific data stored on the WLCG, the PanDA system currently employs a simple heuristic: in principle, it assigns computing jobs to the site that already hosts the required input data. 
In contrast, the Rucio system not only prepares data for jobs at the right time and location, but also considers overall system load and automatically re-balances data.  
However, by adhering to these oversimplified principles and overlooking exceptional cases, it remains unclear how closely the two approaches together approximate true optimality. Moreover, the two systems do not share the same design principles or attempt to optimize the same set of metrics, leaving the overall efficiency of the combined system uncertain. For example, minimizing input data movement reduces network traffic but can overload compute resources at a single site, thereby degrading job throughput and shifting failure patterns from the network to the compute infrastructure.
The critical challenge for system design and evaluation is to acquire sufficient dynamic system information to guide both data placement and job allocation decisions in real time. Yet such information is not always available when needed, or not available at all.


\vspace{-0.2cm}
\subsection{System Performance and Resilience Pitfalls}
\vspace{-0.4cm}
\begin{figure}[h!]
    \includegraphics[width=1.03\columnwidth]{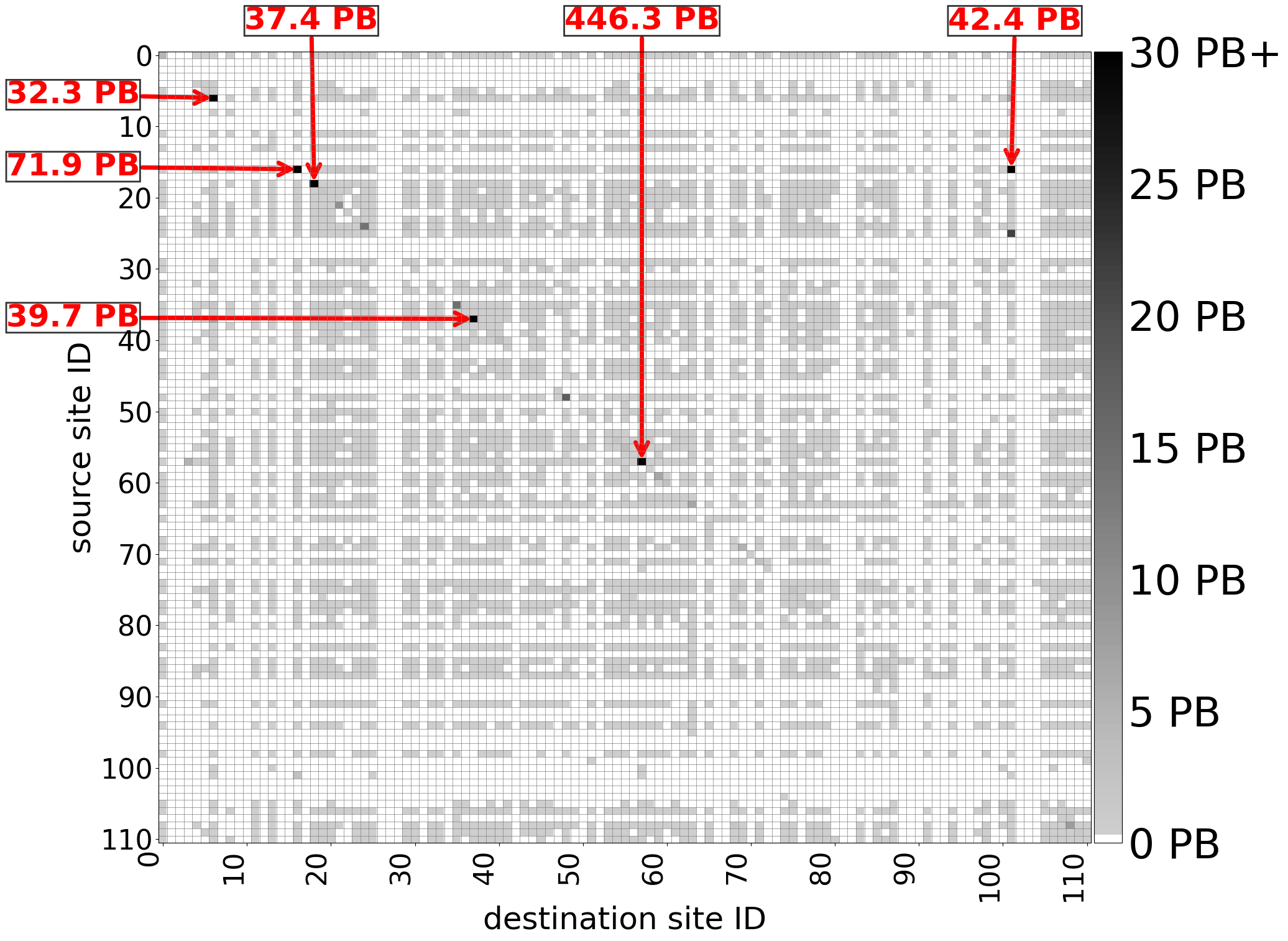}
    \caption{Example file transfer pattern among computing sites in terms of total transferred file size during 3 months, May to July 2025.
    }
    \label{fig:heatmap}
    \vspace*{-0.2cm}
\end{figure} 

Figure~\ref{fig:heatmap} shows the file transfer pattern among computing sites during the selected 92-days time period from 05/01/2025 to 07/31/2025. Each colored cell ($i$, $j$) in this heatmap indicates the total size of all files transferred from source site $i$ (y-axis) to destination site $j$ (x-axis). We observe that many $i$ to $j$ connections involve file transfers, particularly remote transfers ($i$ != $j$), indicating that the Rucio system manages file transfers in a global and intelligent manner. Among the total data volume of 957.98 petabyte (PB) transferred during this period, 737.85 PB correspond to local transfers (diagonal cells), indicating PanDA’s principle of assigning jobs to sites that already host the required input data. However, the file transfer pattern is extremely imbalanced: while the average total file size across all site pairs is 77.75 terabytes (TB) and the geometric mean is only 1.11 TB, numerous outliers exceed 30 PB. Notable examples include 32.2 PB at location (6, 6) (NY, USA, Tier-1), 71.9 PB at location (16, 16) (CERN, Tier-0), 37.4 PB at location (18, 18) (Switzerland, Tier-2), 39.7 PB at location (37, 37) (France, Tier-2), and 446.3 PB at location (North Europe, Tier-1).

Of the 111 sites that recorded file transfers during the observation period, the 102nd site is labeled as $unknown$, aggregating all transfers with either an unidentified source or destination. This classification explains another outlier of 42.4 PB at location (16, 101), representing transfers from CERN to an unknown site.



Under the two main design principles of both PanDA system and Rucio system, the WLCG supports massive data movement across the grid, but with significant spatial and temporal imbalance. While each system achieves its separate design goals, these transfer patterns expose system vulnerability and increase the likelihood of errors at network and storage hot spots. For improved resilience and resource utilization, further investigation is needed to assess the severity of these issues and to guide the development of strategies for system improvement.


\vspace{-0.2cm}
\section{Methodology}
\label{sec:dataanalysis}
\vspace{-0.1cm}
Our methodology provides a structured way to study how PanDA and Rucio interact within the ATLAS distributed computing environment. It integrates metadata from both systems to reconstruct job--transfer relationships despite missing direct identifiers. By applying scalable matching algorithms and correlation techniques, we transform raw logs into interpretable insights that support performance characterization and anomaly detection.

\vspace{-0.2cm}
\subsection{Analysis Workflow}
\vspace{-0.4cm}
\begin{figure}[h!]
    \includegraphics[width=1\linewidth]{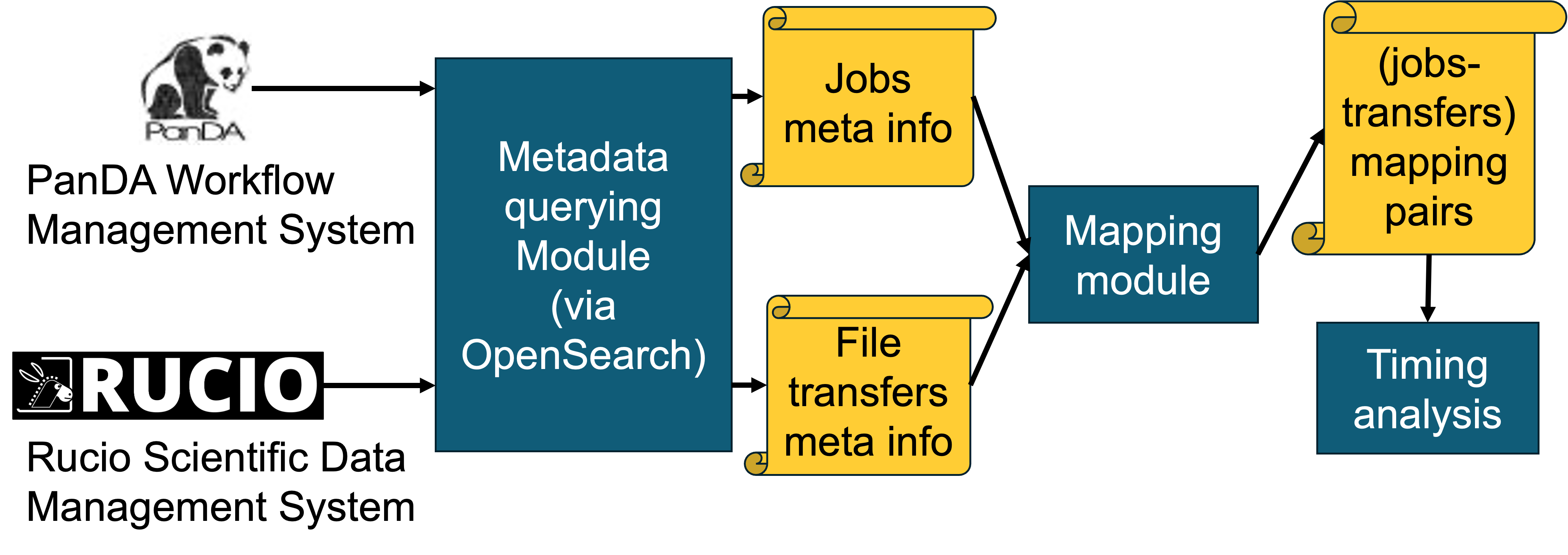}
    \caption{Analysis Workflow
    }
    \label{fig:workflow}
    \vspace*{-0.2cm}
\end{figure}

Figure~\ref{fig:workflow} illustrates an overview of the analysis workflow for this study. The OpenSearch~\cite{opensearch} framework-based querying module retrieves job metadata from PanDA and file and transfer-event metadata from Rucio. These metadata describe real-time events individually and, in principle, provide sufficient information for downstream analysis tasks to characterize the distributed system. However, because not every file transfer event is triggered by a computing job, and job identifiers are not recorded, the absence of a direct temporal mapping between jobs and transfer events obscures a fine-grained understanding of real-time behavior. To address this, the mapping module incorporates file-level metadata to build such connections, producing pairs of jobs and their associated transfer events. These mappings enable a more detailed assessment of system performance and support the exploration of co-optimization opportunities between PanDA and Rucio.





\vspace{-0.2cm}
\subsection{The Exact Mapping Module}
\setlength{\textfloatsep}{1pt plus 2pt minus 2pt} 
\begin{tightfloat} 
\begin{algorithm}[h!]
\caption{Mapping Jobs and File Transfers metadata}  
\label{alg:algo}
\Input{
  Job set:                  $\mathbf{J} = \{J_j \mid 0 \le j < |\mathbf{J}|\,\}$, \\%
  \quad File set:           $\mathbf{F} = \{F_f \mid 0 \le f < |\mathbf{F}|\,\}$, \\%
  \quad Transfer event set: $\mathbf{T} = \{T_t \mid 0 \le t < |\mathbf{T}|\,\}$%
}
\Output{
  mapping set $\mathbf{M} =\{(J_j,T_j)\mid T_j\subseteq\mathbf T,\;|T_j|=N_j,\;N_j\in\mathbb N\}$
}

\Let{$\mathbf{M} \gets \emptyset$}\;
\For{$j \leftarrow 0$ \KwTo $|\mathbf{J}|-1$}{

    \textbf{let} subset $\mathbf F'_j$ = {$\{\,F_f\in\mathbf F\mid 
        F_f.\mathit{pandaid}=J_j.\mathit{pandaid}
        \;\wedge\;
        F_f.\mathit{jeditaskid}=J_j.\mathit{jeditaskid}\}$}\;
           
    \textbf{let} subset $\mathbf T'_j \gets \emptyset$\;
    
      \ForEach{$F_f\in\mathbf F'_j$}{
        $\mathbf T'_j \leftarrow \mathbf T'_j
          \;\cup\;
          \{\,T_t\in\mathbf T\mid$
        \Indp
          \hspace{1.5em}$T_t.\mathit{lfn}=F_f.\mathit{lfn}$
          $\;\wedge\;
           T_t.\mathit{dataset}=F_f.\mathit{dataset}$
          $\;\wedge\;
           T_t.\mathit{proddblock}=F_f.\mathit{proddblock}$
          $\;\wedge\;
           T_t.\mathit{scope}=F_f.\mathit{scope}$
          $\;\wedge\;
           T_t.\mathit{file\_size}=F_f.\mathit{file\_size}$
        \Indm
        $\}$\;
      }    
      
    $\mathbf T'_j \leftarrow \{\,T_t\in\mathbf T'_j \mid (T_t.\mathit{starttime}<J_j.\mathit{endtime}) \;\wedge\; \big(S_j = J_j.\mathit{ninputfilebytes}\;\lor\;S_j = J_j.\mathit{noutputfilebytes}\big) \;\wedge\; \big((T_t.\mathit{is\_download} \wedge T_t.\mathit{destination\_site}=J_j.\mathit{computing\_site}) \;\lor\; (T_t.\mathit{is\_upload} \wedge T_t.\mathit{source\_site}=J_j.\mathit{computing\_site})\big)\,\}$, \quad $S_j = \sum_{T_t\in\mathbf T'_j} T_t.\mathit{file\_size}$\;
        
    \textbf{add} $\{J_j, \mathbf T'_j \}$  \textbf{to} $\mathbf{M}$ \;
}
\Return{$\mathbf{M}$}\;
\end{algorithm}
\end{tightfloat}

Algorithm~\ref{alg:algo} describes the method to construct relations between jobs and file transfer events. Given a set of jobs $\mathbf{J}$, a set of files $\mathbf{F}$, and a set of file transfer events $\mathbf{T}$, this algorithm returns a set of mappings $\mathbf{M}$, where each element is a pair consisting of one job $J_j$ and a set of matched transfer events~$T_j$ associated with the job. Since transfer events do not directly contain the job identifier $\mathit{pandaid}$, the method leverages shared attributes of file records to bridge the relationship between jobs and transfers. For each job $J_j$, there exists a subset of files $\mathbf F'_j$ where each file contains the common task identifier $\mathit{jeditaskid}$ and job identifier $\mathit{pandaid}$ corresponding to $J_j$. This subset $\mathbf F'_j$ is then used to filter the transfer events, producing a subset $\mathbf T'_j$ that contains only those transfer events potentially associated with the files in $\mathbf F'_j$. To refine this mapping, the algorithm applies additional matching using shared attributes between files and transfers, including the logical file name $\mathit{lfn}$, dataset name $\mathit{dataset}$, block-level data identifier $\mathit{proddblock}$, and file size $\mathit{file\_size}$. This process generates transfer candidates for each job. 

A final filtering step is then applied to $\mathbf T'_j$, retaining only transfers that satisfy all of the following conditions: (1) The transfer started before the end time of the job $J_j$, (2) The sum of the file sizes of the associated files equals either the total required input file size or the output file size of the job $J_j$, and (3) For download transfers, the transfer's destination site matches the computing site of $J_j$; for upload transfers, the transfer's source site matches the computing site of $J_j$. For simplicity, this filtering step treats $\mathbf T'_j$ as a whole set rather than solving the underlying NP-hard problem of subset selection with a combinatorial method. In practice, however, the number of candidate transfers per job is typically small, making this approach computationally feasible.

Note that since all metadata are time-series data continuously generated by the real systems, we pre-selected the job set, file set, and transfer set within a common time window. This reduces input size and avoids unnecessary searching overhead. The selected period should be no shorter than the end-to-end lifetime of the jobs of interest, typically spanning days or more, since the query module only reports jobs that are completed before the end of the interval, excluding all jobs still running at that time. For each job, the lifetime is defined as the interval from creation to completion. Within this period, the queuing time is the duration from creation until the recorded start of execution, while the wall time is the execution period from start until completion. Thus, the common time period pre-selection serves as a first, straightforward matching criterion before applying Algorithm~\ref{alg:algo}.

\vspace{-0.2cm}
\subsection{Relaxed Mapping Alternatives}
\vspace{-0.1cm}
Due to the error-prone nature of the metadata retrieval process, the exact mapping algorithm may not always yield a set of matched transfers for every job. To address this limitation, we introduce relaxed matching approaches that increase the likelihood of identifying valid associations.

The first level, relaxed matching approach (\RMone), ignores the file-size checking criterion used in exact matching: \vspace{-0.07cm}\[
\begin{aligned}
S_j = J_j.\mathit{ninputfilebytes} \;\lor\; S_j = J_j.\mathit{noutputfilebytes}
\end{aligned}
\vspace{-0.07cm}\]. This check requires the total file size to match exactly in bytes, which excludes two common cases: (1) when no exact size match exists but a subset of transfers is potentially valid, and (2) when file sizes are not recorded precisely down to the byte level and are therefore rejected by the strict check. Since any advanced algorithm trying to capture these cases would still be approximate, we remove the file-size constraint in \RMone{} for simplicity.

The second level, relaxed matching approach (\RMtwo), further relaxes the site-checking requirement applied in \RMone{}: \vspace{-0.07cm}\[
\begin{aligned}
&(T_t.\mathit{is\_download} \land T_t.\mathit{destination\_site}=J_j.\mathit{computing\_site})\\
&\quad{}\lor{}
 (T_t.\mathit{is\_upload} \land T_t.\mathit{source\_site}=J_j.\mathit{computing\_site})
\end{aligned}
\vspace{-0.07cm}
\]. In many cases, either the source site or destination site is recorded as $unknown$ or with an invalid name. Instead of discarding such transfers as mismatched, \RMtwo{} retains them, recognizing that these site labels may be incorrectly recorded in the metadata while still corresponding to valid matches in the real system. In practice, many of the matches identified through \RMone or \RMtwo show strong evidential validity, and in some \RMtwo{} cases the missing or incorrect site information can be inferred. Additional case studies illustrating these situations are presented in Section~\ref{sec:experiments}.

\vspace{-0.2cm}
\section{Analysis Results}
\label{sec:experiments}
\vspace{-0.1cm}
Using the developed framework for linking PanDA jobs and Rucio file transfers via exact and relaxed strategies, we conduct an analysis of large-scale ATLAS metadata over an 8-day period from 04/01/2025 to 04/09/2025 for the following results. The analysis provides a quantitative view of how job execution and data movement interact, highlighting both typical patterns and anomalous behaviors. We summarize overall matching statistics, compare the outcomes of exact versus relaxed strategies, present detailed case studies that illustrate the systemic inefficiencies and resilience challenges uncovered by our approach, and discuss some limitations of the approach.

\begin{figure*}[t]
  \centering
  \includegraphics[width=1.01\textwidth]{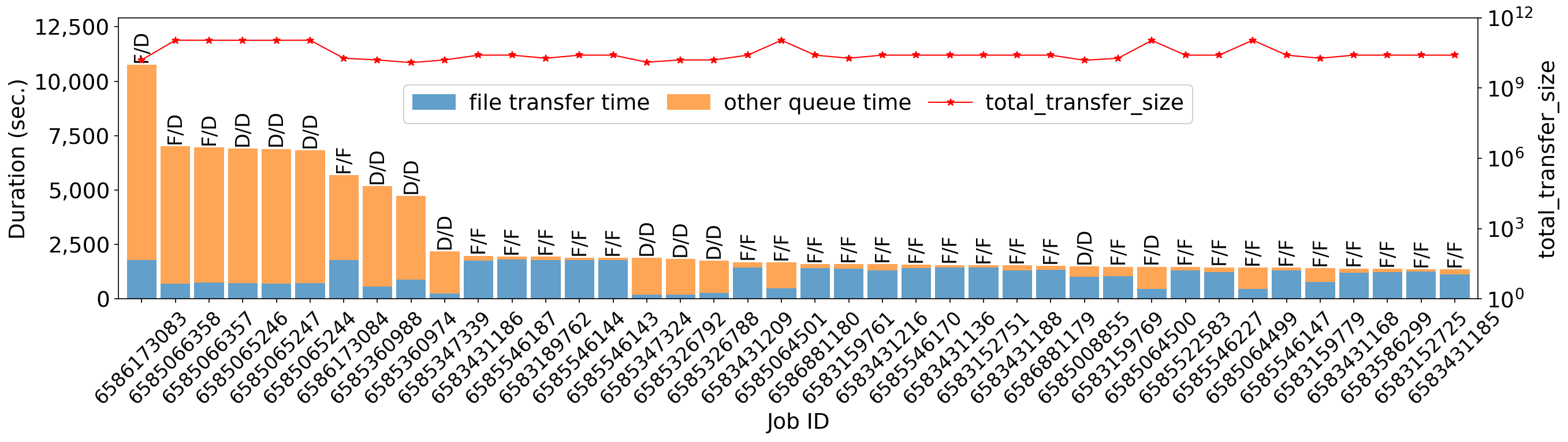}
  \vspace{-0.5cm}
  \caption{Top 40 jobs with local transfers that last for more than 10\% of the job queuing time.}
  \label{fig:job_duration_local}
  \vspace{-0.4cm}
\end{figure*}

\vspace{-0.2cm}
\subsection{Summary of Exact Matching}
\vspace{-0.1cm}
Over the 8-day study period, we collected 966,453 user jobs and 6,784,936 file-level transfer events globally. Among these transfers, 1,585,229 of them contained a valid $jeditaskid$. Using the exact matching, 30,380 transfers with $jeditaskid$ were successfully linked to jobs, which is 1.92\% of all transfers with a $jedistaskid$. Similarly, among nearly one million user jobs, the exact matching identified 7,907 of them successfully, which is 0.82\%. For the matched job-transfers pairs, we observe that the average transfer time occurring during job queuing time was 8.43\%, with a geometric mean of 1.942\%. In this analysis, file transfer time is defined as the cumulative duration during the job's queuing time phase in which at least one associated file was actively transferring. This result is consistent with the design principle of PanDA system, in which jobs are primarily assigned to sites where the required input data are already located. Thus, transfer time does not typically constitute a critical bottleneck for queuing latency. However, we also observe that the matched transfer timelines are not always aligned across jobs, prompting a closer examination of anomalies, which we present in Section~\ref{casestudy}.

\begin{table}[h]
\vspace{0.1cm}
\caption{Breakdown of Exact Matched Transfers}
\centering
\resizebox{\columnwidth}{!}{%
\begin{tabular}{@{}l r r r@{}}
\toprule
\textbf{Transfer activity type} & \textbf{Matched count} & \textbf{Total count} & \textbf{Percentage} \\
\midrule
Analysis Download & 14{,}811 & 176{,}694 & 8.38\% \\
Analysis Upload & 2{,}919 & 3{,}059 & 95.42\% \\
Analysis Download Direct IO & 12{,}650 & 548{,}712 & 2.31\% \\
Production Upload & 0 & 824{,}963 & 0\% \\
Production Download & 0 & 31{,}801 & 0\% \\
\midrule
\textbf{Total} & \textbf{30{,}380} & \textbf{1{,}585{,}229} & \textbf{1.92\%} \\
\bottomrule
\end{tabular}%
\label{tab:breakdown}
}
\vspace{0.1cm}
\end{table}

Table~\ref{tab:breakdown} presents the activity breakdown of the transfers matched by the exact strategy. We observe that nearly all transfers that have $jeditaskid$ fall to the following activities: (1) Analysis Download - files are transferred before job execution, (2) Analysis Upload - files are transferred after job completion, (3) Analysis Download Direct IO - file transfers occur in streaming mode and overlap with job execution, and (4) Production Upload and (5) Production Download are for production jobs only. Among these, 95\% of the Analysis Upload transfers can be matched to jobs, since this transfer scheme is relatively straightforward. As for Analysis Download, however, the fraction of queuing time spent on file transfer varies significantly, ranging from nearly zero (e.g. a transfer lasting only one second) to more than 83\%. In some cases, transfers begin after the job start time, which is expected behavior for local transfers. In other cases, transfer durations span across both the job queuing time and execution time, which lead to anomalous operation likely caused by errors.

\vspace{-0.2cm}
\subsection{Summary of Relaxed Matching}
\vspace{-0.1cm}
\begin{table}[h]
\centering
\caption{Matched transfers and matched job counts by matching methods.}
\vspace{0cm}
\label{tab:match-split}

\begin{subtable}{\linewidth}
\centering
\caption{Matched transfers count}
\label{tab:transfers}
\renewcommand{\arraystretch}{1.2}
\setlength{\tabcolsep}{6pt}
\begin{adjustbox}{max width=\linewidth}
\begin{tabular}{l r r r c}
\toprule
\textbf{Matching method} &
\makecell{Local\\transfer} &
\makecell{Remote\\transfer} &
\makecell{Total\\transfer} &
\makecell{Total\\matched \%} \\
\midrule
Exact & 28,579 & 1,801  & 30,380 & 1.92\% \\
\RMone   & 35,065 & 1,817  & 36,882 & 2.33\% \\
\RMtwo  & 36,320 & 24,273 & 60,593 & 3.82\% \\
\bottomrule
\end{tabular}
\end{adjustbox}
\label{tab:rma}
\end{subtable}

\vspace{0.2cm}

\begin{subtable}{\linewidth}
\centering
\caption{Matched job count}
\label{tab:jobs}
\renewcommand{\arraystretch}{1.2}
\setlength{\tabcolsep}{6pt}
\begin{adjustbox}{max width=\linewidth}
\begin{tabular}{l r r r r c}
\toprule
\shortstack[l]{\textbf{Matching}\\\textbf{method}} &
\makecell{Jobs with\\all local\\transfers} &
\makecell{Jobs with\\all remote\\transfers} &
\makecell{Jobs with\\mix\\transfers} &
\makecell{Total\\jobs} &
\makecell{Total\\matched \%} \\
\midrule
Exact & 7,649 & 258  & 0   & 7,907  & 0.82\% \\
\RMone   & 8,763 & 260  & 0   & 9,023  & 0.93\% \\
\RMtwo   & 8,727 & 7,662 & 112 & 16,501 & 1.71\% \\
\bottomrule
\end{tabular}
\end{adjustbox}
\label{tab:rmb}
\end{subtable}
\label{tab:rm}
\vspace{-0cm}
\end{table}

Table~\ref{tab:rm} summarizes the matched transfer counts and matched job counts across the three different matching methods. In Table~\ref{tab:rma}, the exact matching method identifies 94\% of the matched transfers as local transfers. Under \RMone{}, the number of local transfers increases to 35,065 due to the relaxation of file-size checking. With \RMtwo{}, an additional 24,273 remote transfers are identified, since the computing site-checking requirement is also relaxed. In Table~\ref{tab:rmb}, the matched job counts follow a similar pattern, increasing from the exact method to \RMone{} and then \RMtwo{}. For example, among the 8,763 jobs identified by \RMone{} with only local transfers, \RMtwo{} reclassifies some as mixed-transfer jobs due to the discovery of additional remote transfers.

\vspace{-0.2cm}
\subsection{Detailed Analysis of Exact Matching}
\vspace{-0.1cm}
To move beyond aggregated statistics, we examine the matched jobs and transfers in greater detail. While most jobs adhere to PanDA’s data locality principle and exhibit only modest transfer delays, a subset shows disproportionately long transfer times that often coincide with failures. This closer inspection highlights both the strengths of current strategies and the anomalies that undermine efficiency and resilience.

\begin{figure*}[t]
  \centering
  \includegraphics[width=1.01\textwidth]{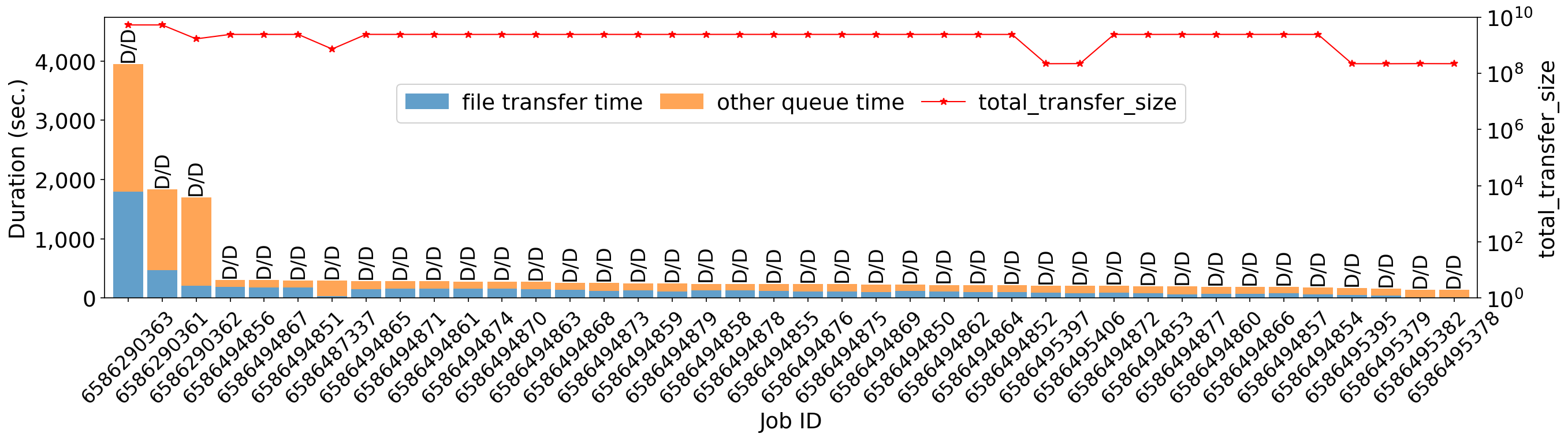}
  \vspace{-0.6cm}
  \caption{Top 40 jobs with remote transfers that last for more than 10\% of the job queuing time.}
  \label{fig:job_duration_remote}
  \vspace{-0.4cm}
\end{figure*}

Figure~\ref{fig:job_duration_local} shows the queuing-time breakdown of the 40 jobs with the longest queuing times, each of which spent at least 10\% of its queuing time on file transfer. All of these jobs involved only local file transfers. Data labels indicate the task and job status, using "D" for completed jobs and "F" for failed jobs. The total size of file transferred for each job is also shown on the secondary y-axis. We found no significant correlation between total transfer size and either queuing time or file transfer time. For the outlier job with the longest queuing time, the absolute file transfer time exceeded 10,000 seconds. Also, many failed jobs exhibited relatively high transfer-time percentage compared to successful jobs, including IDs 6583431186, 6585546187, 6583189762, 6585546144, among others. These extreme cases tend to fail, resulting in unnecessarily long transfer duration and low bandwidth utilization.  Applying the same criteria to jobs with only remote transfers, Figure~\ref{fig:job_duration_remote} shows their queuing-time breakdowns. In contrast to the local cases, jobs with remote transfers exhibit relatively stable transfer-time percentages. On the other hand, extreme local cases have much longer job queuing time than their remote counterparts. This suggests that some individual sites experienced server queuing delays despite using local transfers. These observations reveal that strictly following PanDA’s data-centric job allocation principle does not always yield the best performance. Assigning jobs to sites with local data can lead to heavy site-level queuing delays, whereas assigning them to remote sites, despite requiring additional transfers, may result in shorter overall queuing times. This is because actual transfer performance depends not on peak throughput but on effective usage under current conditions, and transfer-related error patterns may shift when alternative sites are used.



\begin{figure}[H]
  \centering
  \begin{subfigure}[t]{\linewidth}
    \centering
    \includegraphics[width=\linewidth,height=3.2cm,keepaspectratio]{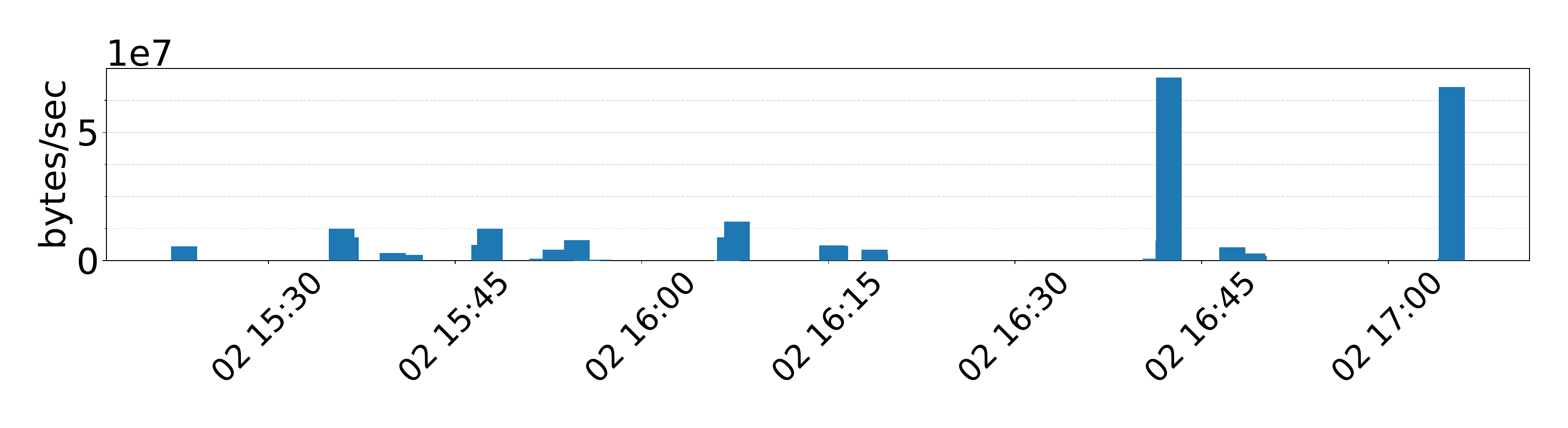}
    \vspace{-0.6cm}
    \caption{From Slovenia, Tier-2 to North Europe, Tier-1}
    \label{fig:throughput_variation_1}
  \end{subfigure}\vspace{0.2em}
\vspace{-0.1cm}
  \begin{subfigure}[t]{\linewidth}
    \centering
    \includegraphics[width=\linewidth,height=3.2cm,keepaspectratio]{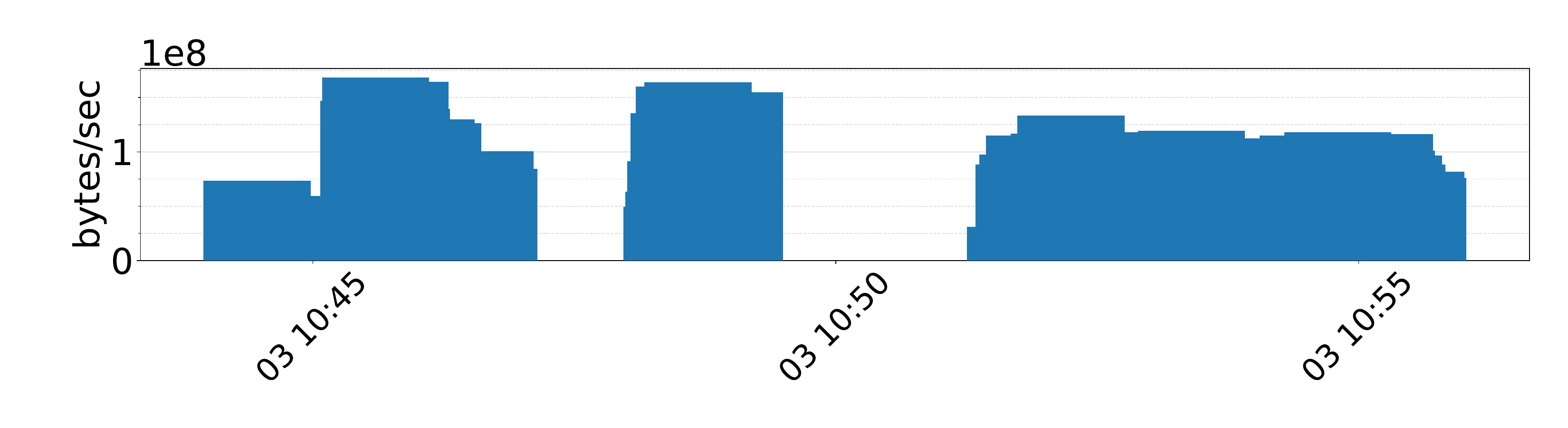}
    \vspace{-0.6cm}
    \caption{From North Europe, Tier-1 to Slovenia, Tier-2}
    \label{fig:throughput_variation_2}
  \end{subfigure}\vspace{0.2em}
\vspace{-0.1cm}
  \begin{subfigure}[t]{\linewidth}
    \centering
    \includegraphics[width=\linewidth,height=3.2cm,keepaspectratio]{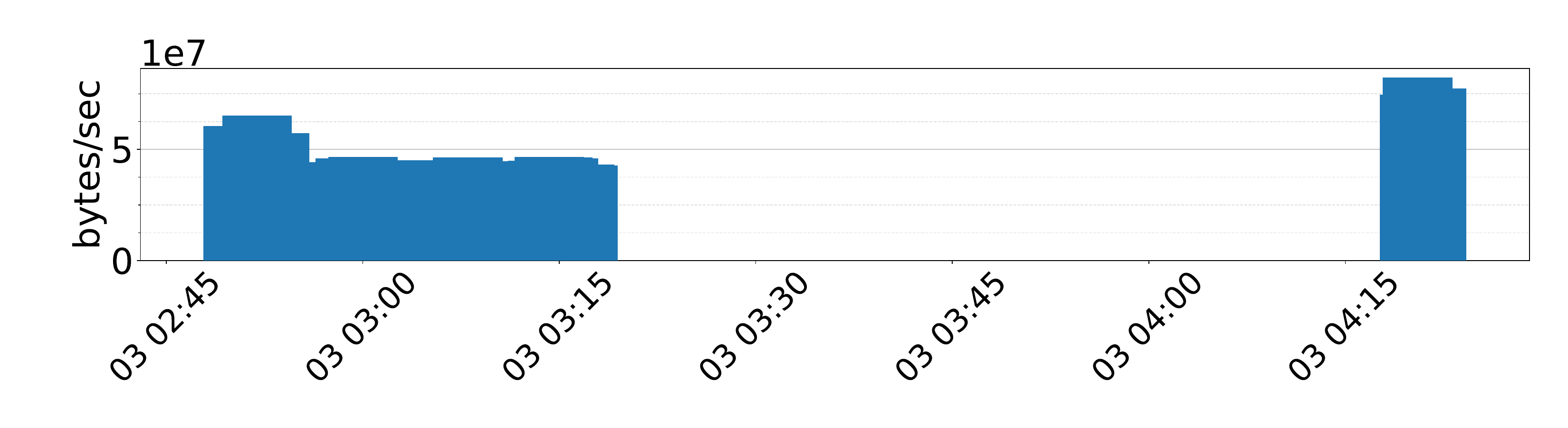}
    \vspace{-0.6cm}
    \caption{From United Kingdom, Tier-1 to United Kingdom, Tier-2}
    \label{fig:throughput_variation_3}
  \end{subfigure}\vspace{0.2em}
\vspace{-0.1cm}
  \begin{subfigure}[t]{\linewidth}
    \centering
    \includegraphics[width=\linewidth,height=3.2cm,keepaspectratio]{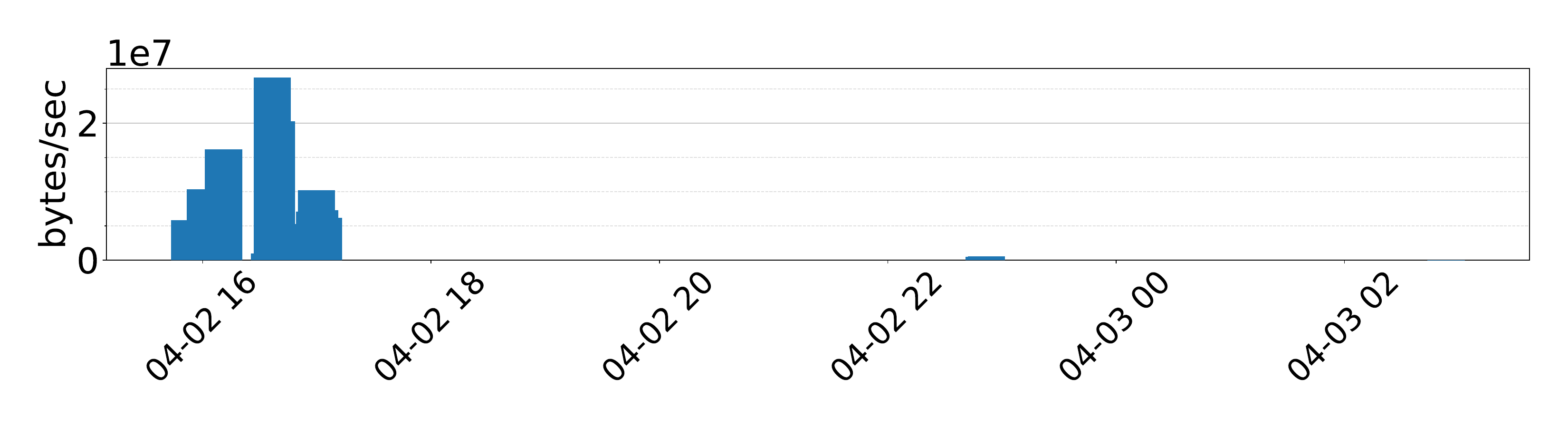}
    \vspace{-0.6cm}
    \caption{From United Kingdom, Tier-2 to United Kingdom, Tier-1}
    \label{fig:throughput_variation_4}
  \end{subfigure}\vspace{0.2em}
\vspace{-0.1cm}
  \begin{subfigure}[t]{\linewidth}
    \centering
    \includegraphics[width=\linewidth,height=3.2cm,keepaspectratio]{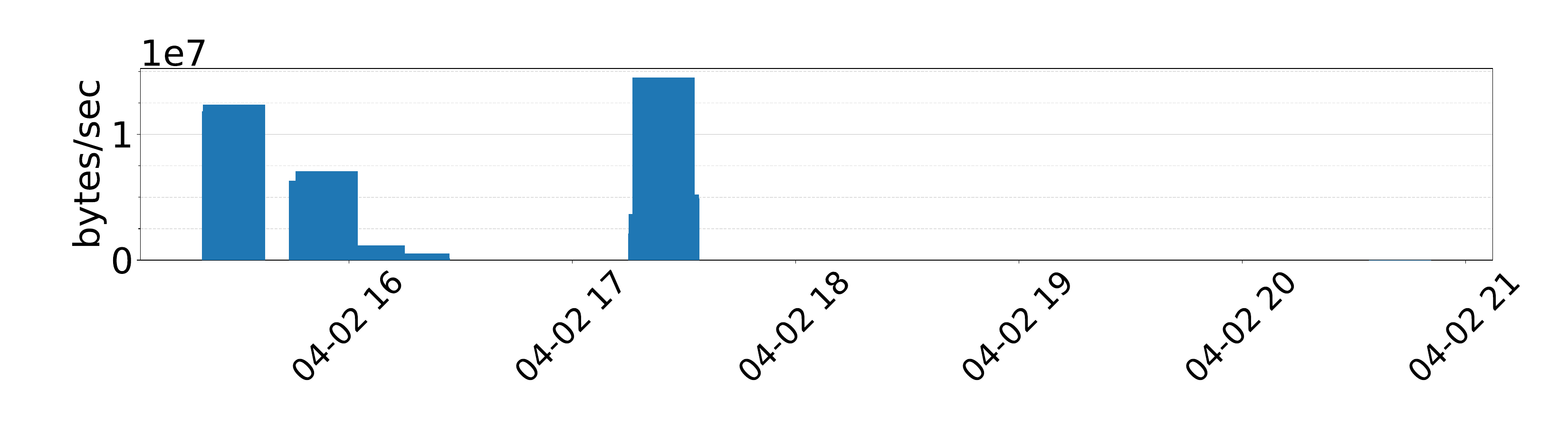}
    \vspace{-0.6cm}
    \caption{From Israel, Tier-2 to Israel, Tier-2 and Tier-3}
    \label{fig:throughput_variation_5}
  \end{subfigure}\vspace{0.2em}
  \vspace{-0.1cm}
  \begin{subfigure}[t]{\linewidth}
    \centering
    \includegraphics[width=\linewidth,height=3.2cm,keepaspectratio]{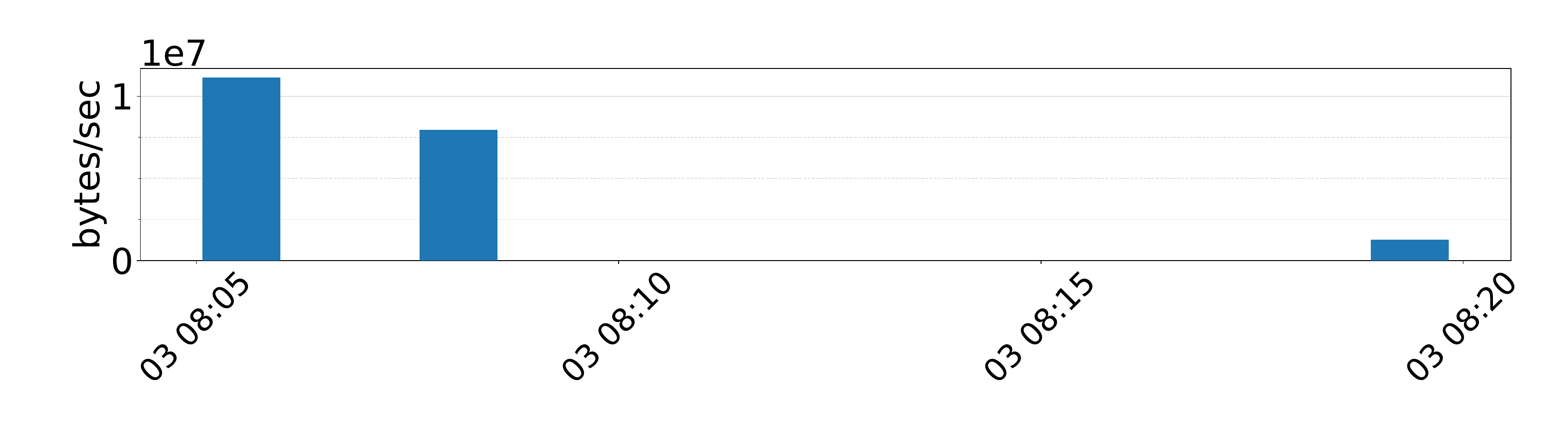}
    \vspace{-0.6cm}
    \caption{From Spain, Tier-1 to Brazil, Tier-2}
    \label{fig:throughput_variation_6}
  \end{subfigure}
    \vspace{0.1cm}
  \caption{Bandwidth usage variation across time at six remote transfer connections.}
  \label{fig:throughput_variation_remote}
  \vspace{-0.5cm}
\end{figure}

Figure~\ref{fig:throughput_variation_remote} provides a detailed illustration of how bandwidth usage varies across several representative remote site-to-site connections. The measurements reveal that transfer rates do not remain steady but instead fluctuate noticeably even within relatively short time intervals. For example, Figure~\ref{fig:throughput_variation_1} shows that the accumulated bandwidth usage of matched transfers during the first few transmission periods remained mostly lower than 10 megabytes per second (MBps), whereas two later periods spiked to over 60 MBps. Interestingly, the actual usage in opposite remote transfer directions is not symmetric. As shown in Figure~\ref{fig:throughput_variation_2}, some transmission periods reached up to 130 MBps. These patterns highlight the variability of cross-site network conditions and indicate that transfer efficiency depends not only on average available bandwidth but also on transient congestion. Such irregularities introduce uncertainty into scheduling and resource management, which in turn can prolong job queuing times and amplify the disparities identified in the matched transfer analysis.


\begin{figure}[h]
  \centering
  \begin{subfigure}[t]{1\linewidth}
    \centering
    \includegraphics[width=\linewidth,height=4cm,keepaspectratio]{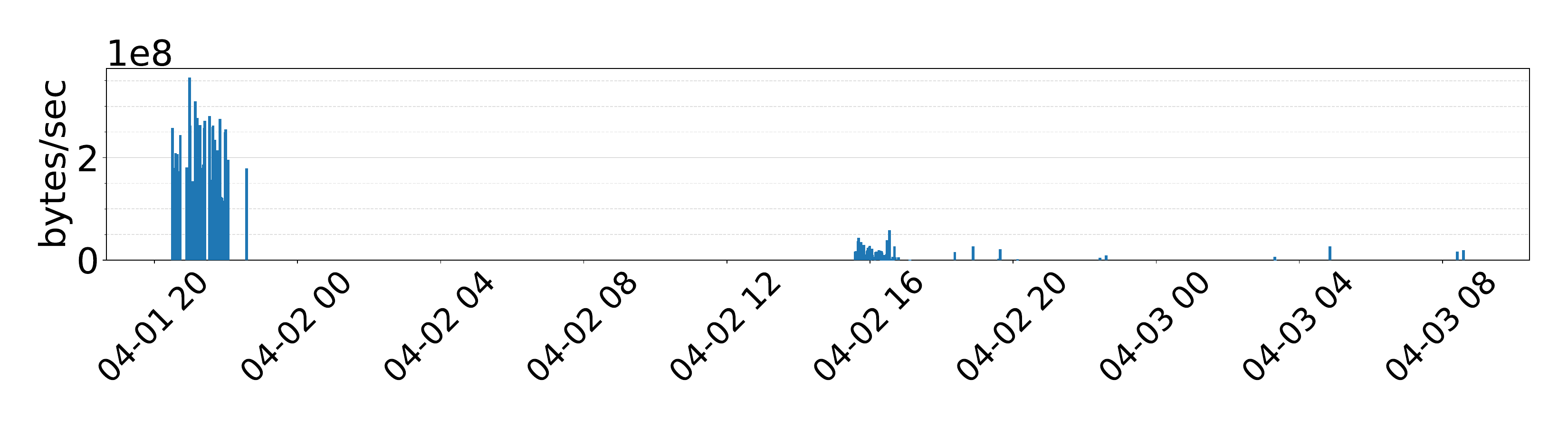}
    \vspace{-0.6cm}
    \caption{Michigan, USA, Tier-2}
    \label{fig:throughput_local_variation_1}
  \end{subfigure}
  \vspace{-0.1cm}
  \begin{subfigure}[t]{1\linewidth}
    \centering
    \includegraphics[width=\linewidth,height=4cm,keepaspectratio]{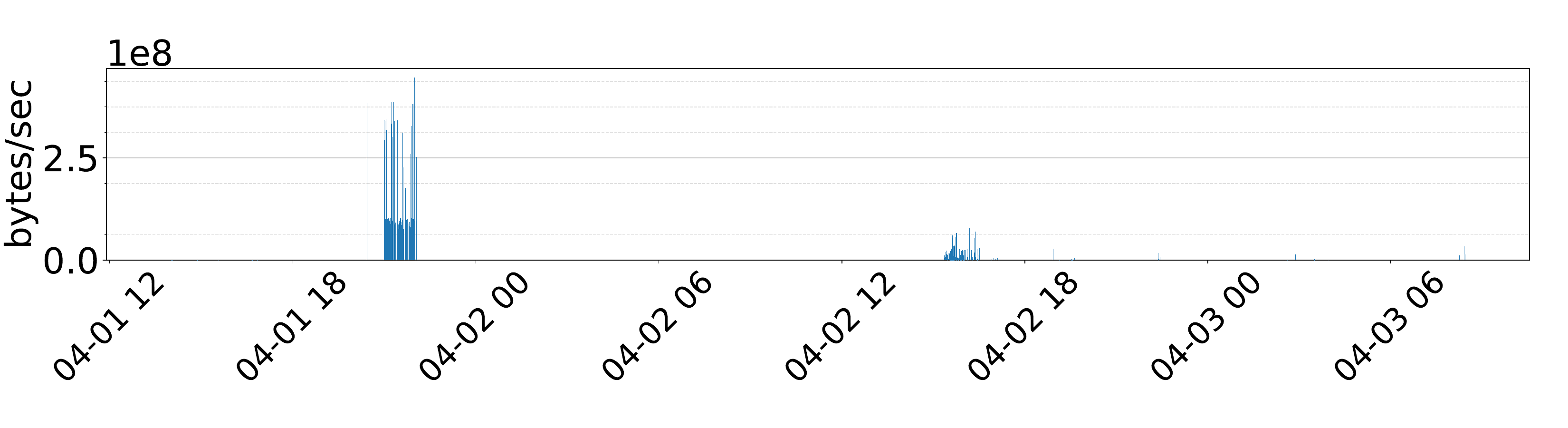}    \vspace{-0.6cm}
    \caption{NY, USA, Tier-1}
    \label{fig:throughput_local_variation_2}
  \end{subfigure}
  \vspace{-0.1cm}
  \begin{subfigure}[t]{1\linewidth}
    \centering
    \includegraphics[width=\linewidth,height=4cm,keepaspectratio]{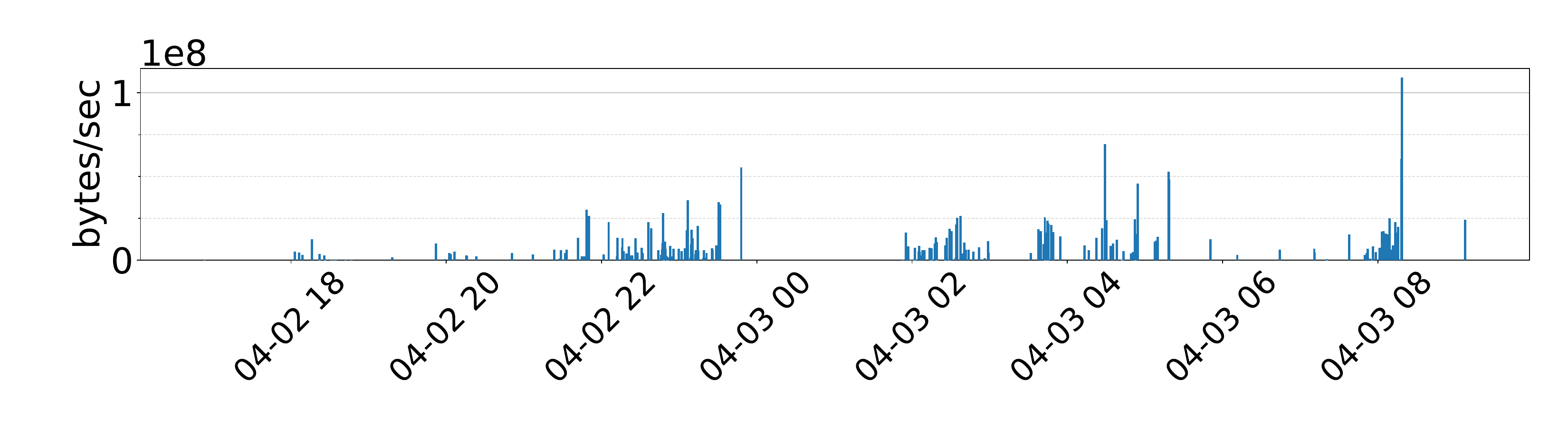}
    \vspace{-0.6cm}
    \caption{Genoa, Italy, Tier-3}
    \label{fig:throughput_local_variation_3}
  \end{subfigure}
  \vspace{-0.1cm}
  \begin{subfigure}[t]{1\linewidth}
    \centering
    \includegraphics[width=\linewidth,height=4cm,keepaspectratio]{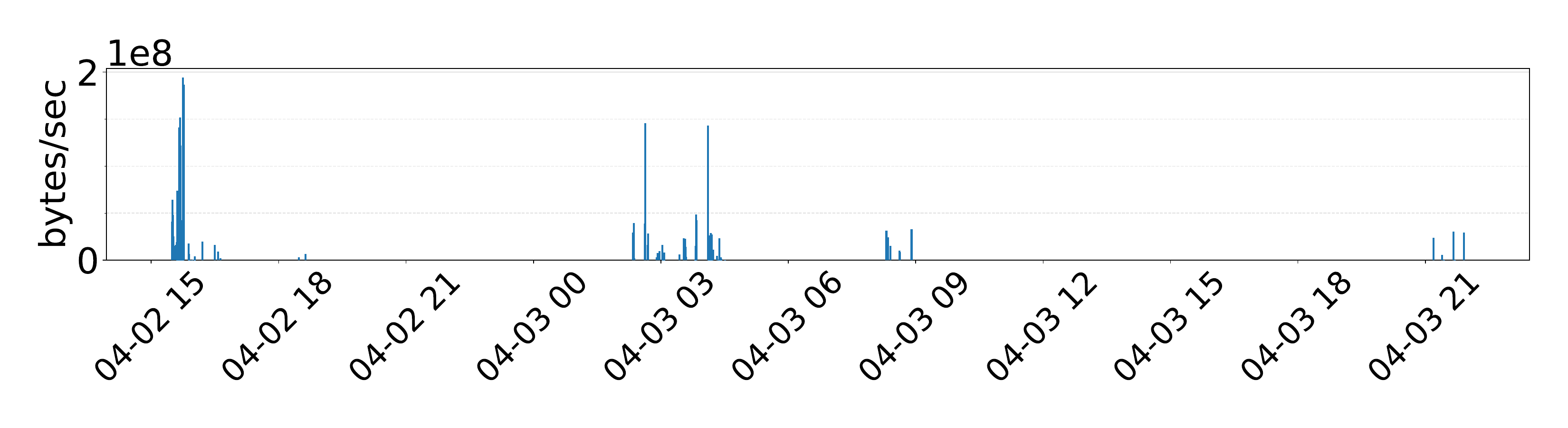}
    \vspace{-0.6cm}
    \caption{Milan, Italy, Tier-2}
    \label{fig:throughput_local_variation_4}
  \end{subfigure}
  \vspace{-0.1cm}
  \begin{subfigure}[t]{1\linewidth}
    \centering
    \includegraphics[width=\linewidth,height=4cm,keepaspectratio]{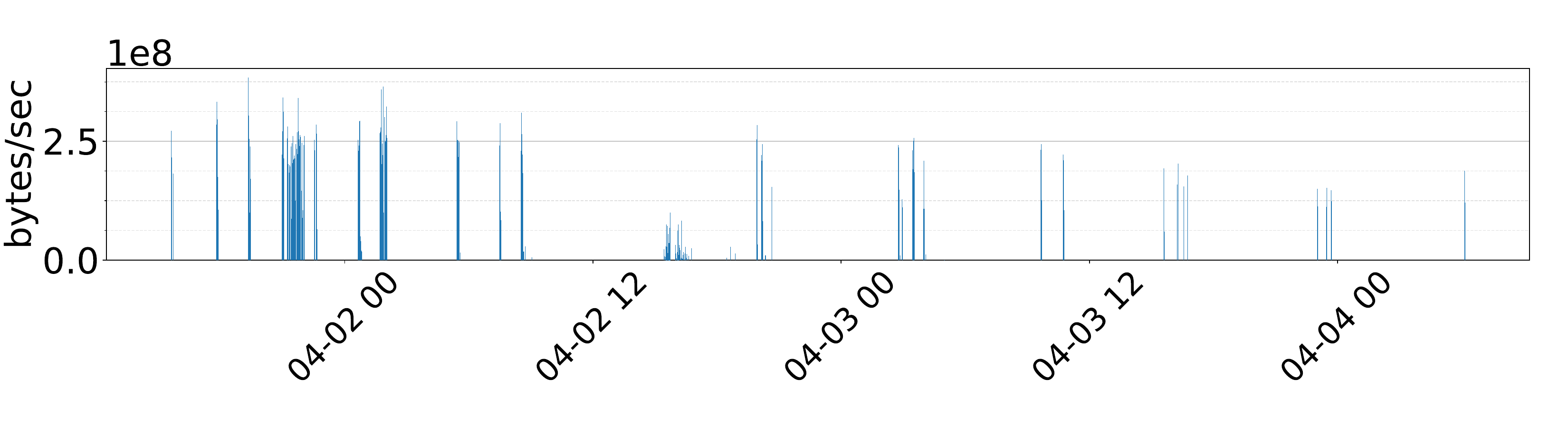}
    \vspace{-0.6cm}
    \caption{Illinois, USA, Tier-2}
    \label{fig:throughput_local_variation_5}
  \end{subfigure}
  \vspace{-0.1cm}
  \begin{subfigure}[t]{1\linewidth}
    \centering
    \includegraphics[width=\linewidth,height=4cm,keepaspectratio]{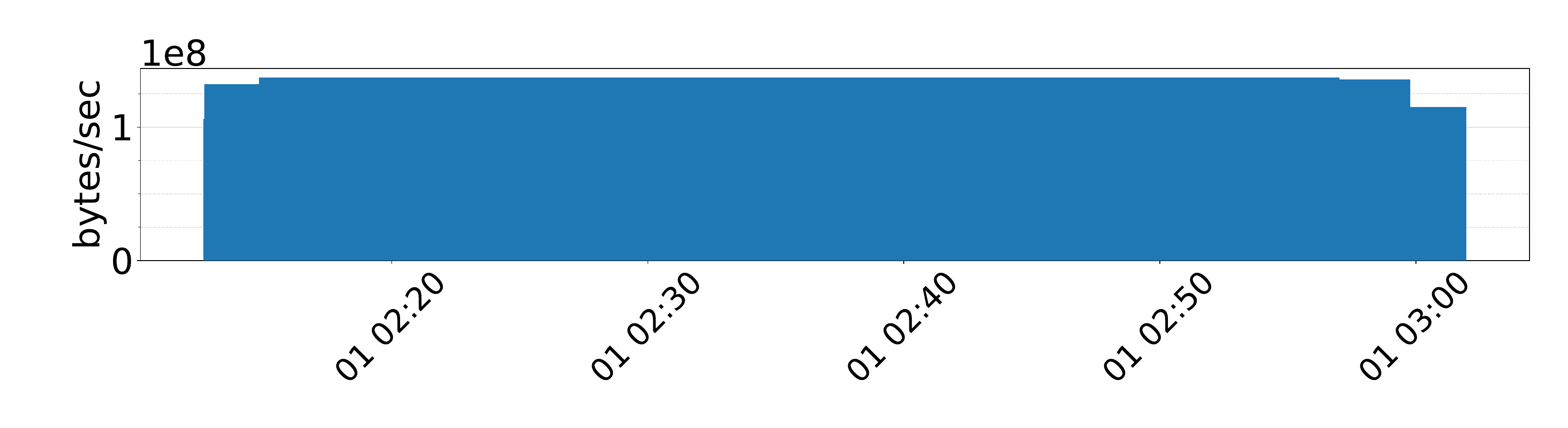}
    \vspace{-0.6cm}
    \caption{Tokyo, Japan, Tier-2}
    \label{fig:throughput_local_variation_6}
  \end{subfigure}

  \caption{Bandwidth usage variation across time at six local sites.}
  \label{fig:throughput_variation_local}
  \vspace{-0.5cm}

\end{figure}

Figure~\ref{fig:throughput_variation_local} presents bandwidth trends at six selected local sites, where transfers are carried out entirely within the same computing facility. Compared to remote transfers, local throughput is generally higher but still exhibits substantial fluctuations over short time intervals. In particular, some sites display intermittent drops that limit effective bandwidth utilization, leading to longer queuing delays even when jobs are scheduled close to the data. For example, during the 9 p.m. period on April 1, as shown in Figure~\ref{fig:throughput_local_variation_2}, bandwidth usage fluctuated heavily, reaching spikes of up to 430 MBps. In contrast, during the 3 p.m. period on April 2, usage remained consistently below 60 MBps. Similar fluctuation patterns are common across many local sites, whereas Figure~\ref{fig:throughput_local_variation_6} shows more stable usage, only due to the shorter observation window and fewer transmission events. These findings reinforce that local transfers are not always optimal in practice and that site-specific bottlenecks can undermine the intended benefits of PanDA’s data locality principle.

Overall, while local transfers generally outperform remote ones, they do not always provide consistent throughput. Site-specific bottlenecks, workload pressure, and scheduling inefficiencies can still reduce the benefits of data locality. As a result, relying solely on local placement may lead to delays, particularly during peak usage periods when contention is high.

\begin{figure}[h]
  \centering
  \vspace*{-0.8cm}
  \hspace*{-0.8cm} 
  \includegraphics[width=0.9\columnwidth, angle=-90]{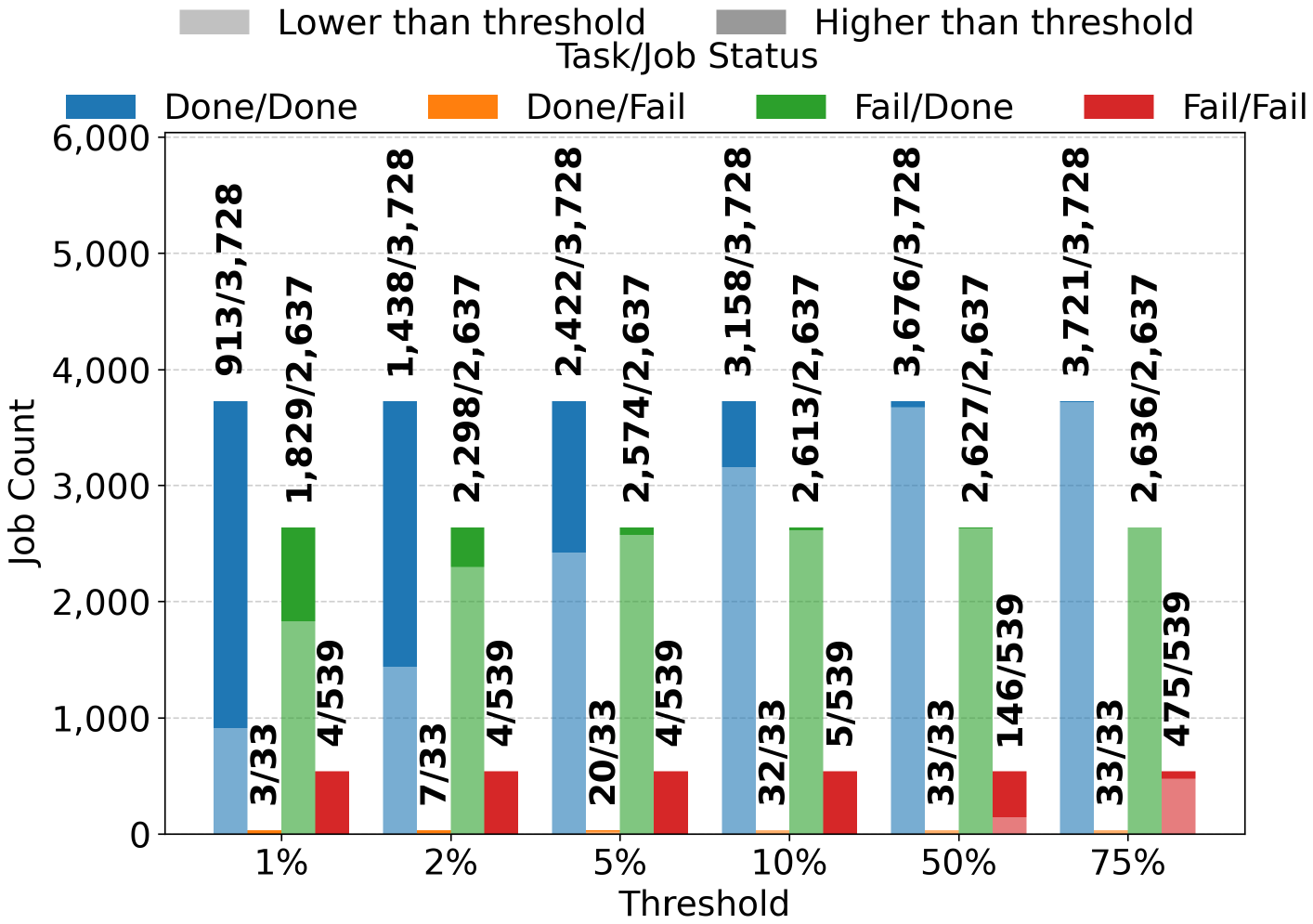}
  \vspace*{-1cm}
  \caption{Job counts of four statuses with various thresholds of transfer-time percentage}
  \vspace*{-0.3cm}
  \label{fig:totals-stacked}
\end{figure}
Figure~\ref{fig:totals-stacked} shows the counts of exactly matched jobs across four status combinations:
(1) the job succeeded within a successful task,
(2) the job failed within a successful task,
(3) the job succeeded within a failed task, and
(4) the job failed within a failed task. For the same set of jobs, we define a threshold \(T\) to divide jobs into two groups based on the percentage of total file-transfer time relative to job queuing time. By varying \(T\), we can observe how many jobs fall into each transfer-time percentage interval. For example, among jobs where both the job and its task were successful, 913 jobs had a transfer-time percentage below \(1\%\), while another 525 jobs (\(1{,}438 - 913\)) fell within the \(1\%\text{--}2\%\) interval. Across all \(7{,}907\) exactly matched jobs, \(6{,}365\) (\(80.5\%\)) were successful (\(3{,}728 + 2{,}637\)).
Even at \(T = 75\%\), there remain \(72\) jobs (\(7 + 0 + 1 + 64\)) with transfer-time percentage greater than \(75\%\), indicating that a small minority of jobs encounter severe performance bottlenecks due to suboptimal file-transfer operations. Notably, most of these extreme cases correspond to failed jobs. These results suggest a potential relationship between high transfer-time percentages and elevated error rates. However, the underlying causes of these correlations require further investigation.

\vspace{-0.3cm}
\subsection{Case Studies}
\label{casestudy}
\vspace{-0.1cm}
In addition to the overview and summary of matching results, we present three case studies that illustrate detailed matching scenarios and findings including (1) a job with extremely long transfer time percentage, (2) a failed job with unexpected transfer-time duration, and (3) a \RMtwo{} matched job involving both reconstructible incomplete metadata and redundant operations.

\begin{figure}[h]
    \vspace*{-2cm}
    \hspace*{-0.8cm} 
    \includegraphics[width=0.88\linewidth, angle=-90]{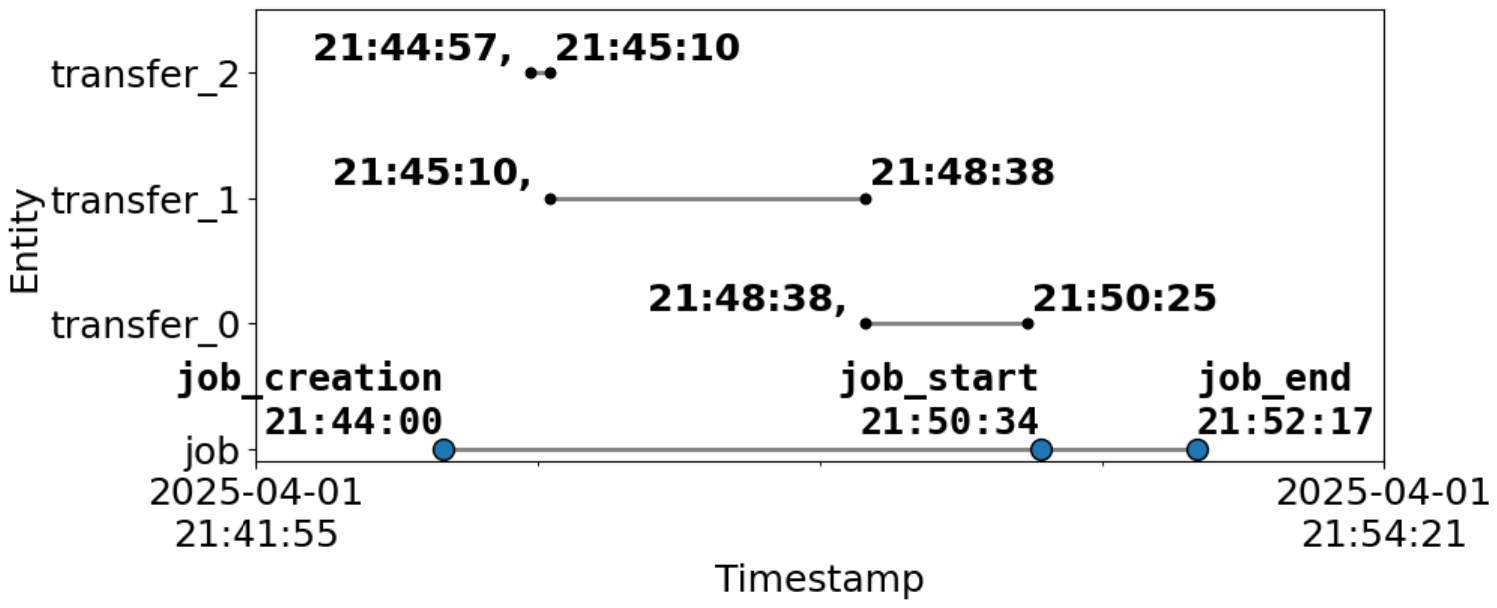}
    \vspace*{-2cm}
    \caption{Timeline of a successful job with local transfers ($pandaid$: 6583770648)
    }
    \label{fig:matched_job_6583770648}
    \vspace*{-0.1cm}
\end{figure}
Figure~\ref{fig:matched_job_6583770648} shows the matching timeline of a successful job and its exact matched local transfers, where 83\% of the job’s queuing time was spent on file transfers. In this example, the total file transfer time was 328 seconds, which constituted the primary source of queuing delay. The three transfer events involved files of sizes 2.1 GB, 4.4 GB, and 4.5 GB (transfers 0, 1, and 2, respectively). Therefore, we infer that local bandwidth was not utilized consistently, as the throughput differed by a factor of approximately \xspeed{17.7} between the first and third transfers. The timeline also shows that the transfers occurred sequentially rather than in parallel. This suggests that the underlying file transfer mechanism doesn't enable parallel file transfers at every site, providing clear evidence of bandwidth underutilization.

\begin{figure}[h]
    \vspace*{-2cm}
    \hspace*{-0.8cm} 
    \includegraphics[width=0.88\linewidth, angle=-90]{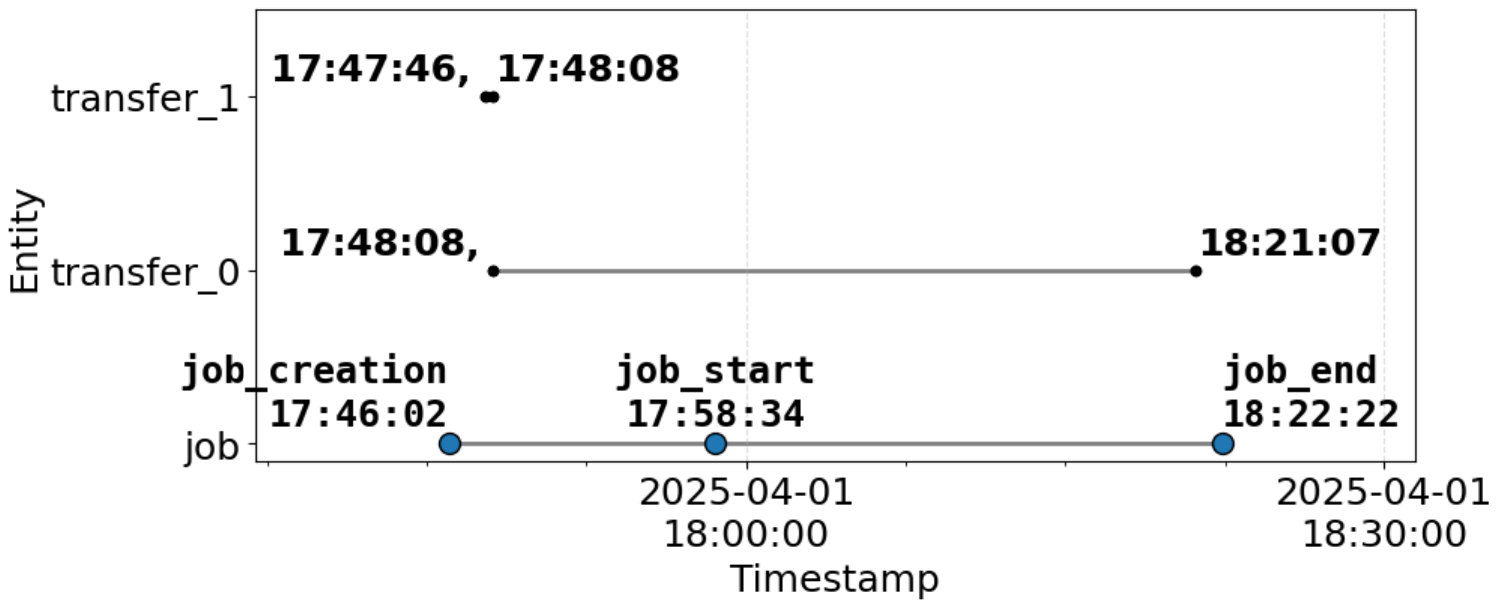}
    \vspace*{-2cm}
    \caption{ Timeline of a failed job with local transfers ($pandaid$: 6583431126)
    }
    \label{fig:matched_job_6583431126}
    \vspace*{-0.4cm}
\end{figure}

Figure~\ref{fig:matched_job_6583431126} shows the matching timeline of a failed job and its exactly matched local transfers. The first transfer (4.6 GB) completed in 22 seconds, while the second transfer (20.5 GB) lasted over 30 minutes, spanning both the job's queuing time and wall time and occupying more than 90\% of the job lifetime. These two transfers exhibited a throughput difference of more than \xspeed{20}. Although this job ultimately failed with execution error code $1305$ and error message: "Non-zero return code from Overlay (1)", it is unclear whether the prolonged transfer directly caused the failure. Nevertheless, even if the job had succeeded, the excessive transfer time would still have been the dominant performance bottleneck. Because this job corresponds to an Analysis Download activity, execution could not begin until the transfer was completed. Therefore, although causality cannot be established, it remains plausible that the lengthy transfer increased the likelihood of failure.

\begin{figure}[h]
    \vspace*{-2cm}
    \hspace*{-0.8cm} 
    \includegraphics[width=0.88\linewidth, angle=-90]{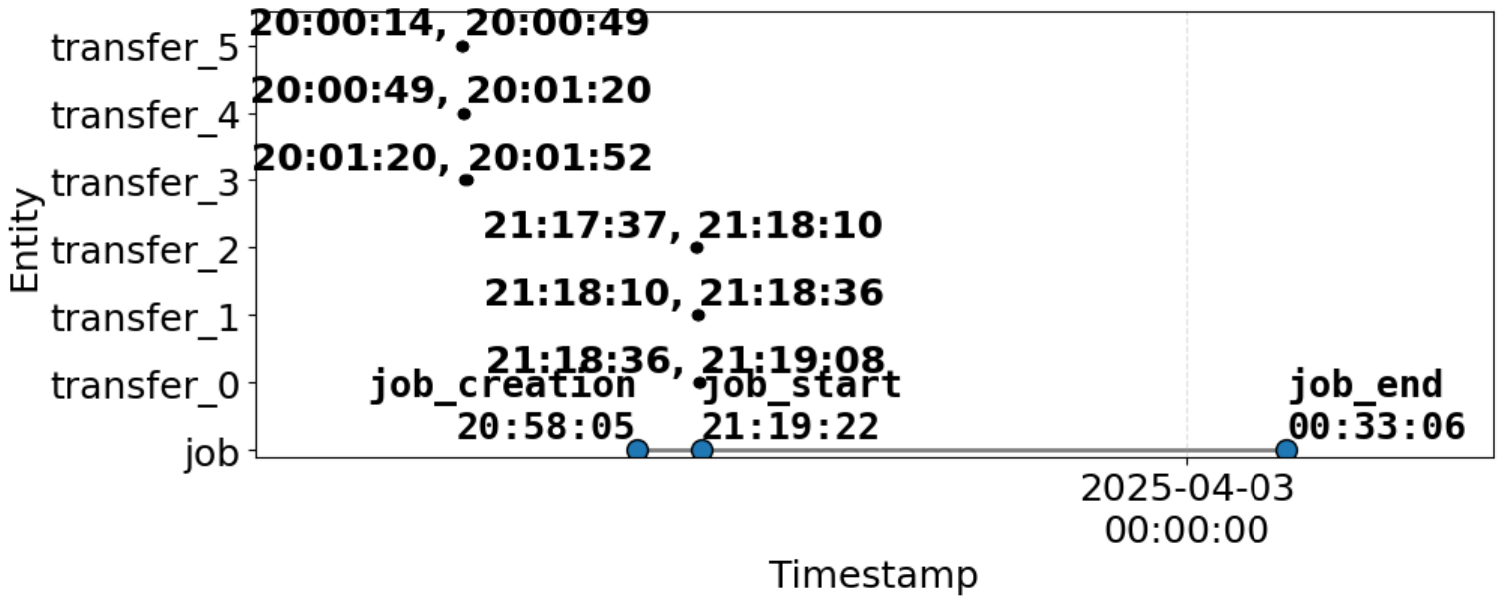}
    \vspace*{-2cm}
    \caption{ Timeline of a \RMtwo{} matched job with abnormal transfers ($pandaid$: 6585617863)
    }
    \label{fig:matched_job_6585617863}
    \vspace*{-0.4cm}
\end{figure}

\begin{table}[h!]
\centering
\caption{Transfer summary of the same job with $pandaid$: 6585617863}
\label{tab:transfer-summary}

\begin{adjustbox}{max width=1\linewidth}
\begin{tabular}{l r r r}
\toprule
\makecell{Field} & \makecell{Transfer 0} & \makecell{Transfer 1} & \makecell{Transfer 2} \\
\midrule
Source Site       & CERN-PROD & CERN-PROD & CERN-PROD \\
Destination Site  & UNKNOWN   & UNKNOWN   & UNKNOWN \\
File Size (Byte)  & 5243410528 & 5243415988 & 5242750540 \\
Activity          & Analysis Download & Analysis Download & Analysis Download \\
Throughput (Byte/s) & 163856579.0 & 201669845.7 & 158871228.5 \\
\bottomrule
\end{tabular}
\end{adjustbox}

\vspace{0.1cm} 

\begin{adjustbox}{max width=1\linewidth}
\begin{tabular}{l r r r}
\toprule
\makecell{Field} & \makecell{Transfer 3} & \makecell{Transfer 4} & \makecell{Transfer 5} \\
\midrule
Source Site       & CERN-PROD & CERN-PROD & CERN-PROD \\
Destination Site  & CERN-PROD & CERN-PROD & CERN-PROD \\
File Size (Byte)  & 5243410528 & 5243415988 & 5242750540 \\
Activity          & Analysis Download & Analysis Download & Analysis Download \\
Throughput (Byte/s) & 163856579.0 & 169142451.2 & 149792872.6 \\
\bottomrule
\end{tabular}
\end{adjustbox}
\vspace*{-0.1cm}
\end{table}

Figure~\ref{fig:matched_job_6585617863} shows a \RMtwo{}-matched and successful job with two sets of identical files transferred. The first set of transfers (0, 1, and 2) are typical local transfer operations that happened immediately before the job's start time, lasting 91 seconds out of the 1,277-second job queuing time (7.1\%). However, the same set of files had already been transferred earlier, prior to the job's creation. From the detailed metadata shown in Table~\ref{tab:transfer-summary}, we observe that the destination site of the first set is recorded as "UNKNOWN" due to a data retrieval error. As a result, the exact matching method cannot identify this case. With \RMtwo{}, however, we can infer that the unknown destination site is actually "CERN-PROD", based on multiple metadata attributes, including the exact files sizes: 5,243,410,528 bytes, 5,243,415,988 bytes, and 5,242,750,540 bytes for transfer pairs (0, 3), (1,4), and (2, 5), respectively. This example demonstrates that relaxed methods not only capture additional possible matches but also help to infer incomplete metadata, effectively converting uncertain cases into exact ones. Therefore, we identified redundant file-transfer patterns, which are in principle avoidable. Many extra examples identified by \RMtwo{} fall into this category.

\vspace{-0.2cm}
\subsection{Limitations}
\vspace{-0.1cm}
Data-management analysis on a globally distributed scientific computing grid such as the WLCG is neither easy nor trivial. The current limitations of a full-scale and comprehensive data movement analysis include (1) the data collection is not only time-consuming but also error-prone, resulting in raw data of uncertain quality for analysis, and (2) multiple metadata information sources complicate the information integration process due to uncoordinated format designs, and (3) even if any such system had unified meta information for both computing job and data transfer, it would not have backward compatibility and still need to provide strong error-free guarantee. Another limitation is the volume of metadata imposes the need for efficient computing for scalability. Instead of developing advanced, efficient analysis methods that try to accommodate the current data quality, any future systematic and scalable analysis designs, such as parallelization, will be especially valuable once data quality improves.

\vspace{-0.2cm}
\section{Related Works}
\label{sec:related}
\vspace{-0.1cm}
Workload and data-management co-optimization at LHC scale has long been a community goal. 
The latest PanDA~\cite{panda} overview explicitly highlights tight integration with the Rucio data-management system as a design principle to sustain ATLAS throughput at scale. 
However, it provides limited empirical evidence about end-to-end effects when both systems operate concurrently across heterogeneous sites. 
Rucio’s reference paper~\cite{rucio}, in turn, introduces the core concepts and mechanisms such as dataset identifiers (DIDs), replication rules, deletion policies, and integration with transfer services that structure bulk data movement and placement across the WLCG. 
In contrast, our study links PanDA job executions with Rucio transfer events at file granularity, exposing coordination gaps that result in redundant transfers, protracted staging delays, bandwidth underutilization, and site-level imbalances that arise despite each system meeting its local objectives.

A complementary line of work targets active orchestration between workload and data systems. 
The intelligent Data Delivery Service (iDDS)~\cite{guan2021intelligent} decouples pre-processing and delivery from execution, orchestrating PanDA and Rucio (e.g., the Data Carousel) to ensure fine-grained, pre-staged data availability and to reduce “long tails” in ATLAS production. 
ServiceX~\cite{choi2021towards} provides another lever by enabling on-demand, columnar extraction and transformation for near-interactive analysis, thereby reducing bytes in flight to analysis facilities. 
Our results are synergistic: whereas iDDS and ServiceX reshape delivery, we diagnose where and when PanDA–Rucio coordination under-delivers, so that pre-staging, replication, or selective delivery policies can be applied where they are most impactful.

Recent facility-scale studies underscore the need for cross-layer correlation. 
For example, the NERSC case study~\cite{giannakou2024understanding} analyzes three years of DTN and border traffic, finding a shift toward medium-sized transfers, DTN over-utilization, and up to 30\% throughput loss during contention. 
It recommends programmable networks, QoS management, and correlation of network and workflow telemetry across an Integrated Research Infrastructure. 
Our analysis performs precisely such cross-system correlation within a production HEP workflow by reconstructing job–transfer relationships. 
This allows us to quantify staging’s impact on job outcomes and to reveal systemic imbalances across sites and timescales. 

Finally, community-wide tests and demonstrations continue to evolve the transfer substrate. 
The WLCG/DOMA Data Challenge 2024 exercised Rucio/FTS at HL-LHC fractions to optimize configurations and expose scaling limits, while SC’22’s SENSE–Rucio–FTS interoperation demonstrated how programmable networks can be surfaced to data-management frameworks for deterministic paths and QoS \cite{lehman2022data}. 
Our results complement these efforts by pinpointing when transfer inefficiencies are caused by workflow–data miscoordination, suggesting that future iterations of challenges and demonstrations incorporate job-level provenance and correlation to target end-to-end performance rather than transfer throughput alone.

Additionally, our analysis provides missing information for modeling distributed grid systems, not only from the job-execution perspective \cite{kilic2025towards,park2024ai} but also from the data-movement perspective. 
This dual view can inform the optimization of both job scheduling and data allocation in Grid Computing \cite{feng2025alternative}.

\vspace*{-0.4cm}

\vspace{-0.2cm}
\section{Conclusion}
\label{sec:conclusion}
\vspace{-0.1cm}
In this work, we presented a systematic analysis of workflow-data interactions within ATLAS, focusing on the coordination between PanDA and Rucio. By developing a fine-grained file-level matching framework, we were able to link approximately 0.8--1.7\% of jobs and 1.9--3.8\% of transfers, enabling visibility into systemic inefficiencies. The analysis revealed imbalances in transfer patterns across sites and timescales, as well as anomalous jobs with extreme transfer-time percentages (>75\%) strongly correlated with failures. Case studies highlighted specific inefficiencies, including bandwidth underutilization, redundant transfers, and incomplete metadata, underscoring how independent optimizations by PanDA and Rucio can inadvertently degrade end-to-end performance.

These findings emphasize the need for tighter co-optimization of workload scheduling and data placement strategies. Future efforts should focus on automating anomaly detection based on transfer-time thresholds, improving metadata completeness and consistency, and developing adaptive strategies where PanDA and Rucio share performance awareness to jointly balance load and data locality. Such improvements will enhance efficiency, reduce wasted resources, and strengthen the resilience of distributed computing workflows at ATLAS and other large-scale scientific experiments.

\section*{Acknowledgments}
\vspace{-0.1cm}
This material is based on work supported by the U.S. Department of Energy, Office of Science, Office of Advanced Scientific Computing Research under Award Number DE-SC-0012704. This work was done in collaboration with the distributed computing research and development program within the ATLAS Collaboration. We thank our ATLAS colleagues for their support, particularly the ATLAS Distributed Computing team's contributions.

\ifarxiv

\else
  \bibliographystyle{ACM-Reference-Format}
  \bibliography{ref}
\fi

\end{document}